\documentclass[12pt, a4paper]{article}
\usepackage[latin1]{inputenc}
\usepackage{graphicx}
\usepackage{amsmath}
\usepackage{amsfonts}
\usepackage{amssymb}
\numberwithin{equation}{section}
\begin{document}
\def\pplogo{\vbox{\kern-\headheight\kern -29pt
\halign{##&##\hfil\cr&{\ppnumber}\cr\rule{0pt}{2.5ex}&\ppdate\cr}}}
\makeatletter
\def\ps@firstpage{\ps@empty \def\@oddhead{\hss\pplogo}%
  \let\@evenhead\@oddhead 
}
\def\maketitle{\par
 \begingroup
 \def\thefootnote{\fnsymbol{footnote}}
 \def\@makefnmark{\hbox{$^{\@thefnmark}$\hss}}
 \if@twocolumn
 \twocolumn[\@maketitle]
 \else \newpage
 \global\@topnum\z@ \@maketitle \fi\thispagestyle{firstpage}\@thanks
 \endgroup
 \setcounter{footnote}{0}
 \let\maketitle\relax
 \let\@maketitle\relax
 \gdef\@thanks{}\gdef\@author{}\gdef\@title{}\let\thanks\relax}
\makeatother
\def\eq#1{Eq. (\ref{eq:#1})}
\newcommand{\nc}{\newcommand}
\def\theequation{\thesection.\arabic{equation}}
\nc{\beq}{\begin{equation}}
\nc{\eeq}{\end{equation}}
\nc{\barray}{\begin{eqnarray}}
\nc{\earray}{\end{eqnarray}}
\nc{\barrayn}{\begin{eqnarray\star}}
\nc{\earrayn}{\end{eqnarray\star}}
\nc{\bcenter}{\begin{center}}
\nc{\ecenter}{\end{center}}
\nc{\ket}[1]{| #1 \rangle}
\nc{\bra}[1]{\langle #1 |}
\nc{\0}{\ket{0}}
\nc{\mc}{\mathcal}
\nc{\etal}{{\em et al}}
\nc{\GeV}{\mbox{GeV}}
\nc{\er}[1]{(\ref{eq:#1})}
\nc{\onehalf}{\frac{1}{2}}
\nc{\partialbar}{\bar{\partial}}
\nc{\psit}{\widetilde{\psi}}
\nc{\Tr}{\mbox{Tr}}
\nc{\tc}{\tilde c}
\nc{\tk}{\tilde K}
\nc{\tv}{\tilde V}
\nc{\CN}{{\mathcal N}}
\newcommand{\ZZ}{\mathbb{Z}}
\newcommand{\CC}{\mathbb{C}}
\newcommand{\PP}{\mathbb{P}}
\newcommand{\RR}{\mathbb{R}}
\nc{\Imtau}{\mbox{Im}\tau}
%

\setcounter{page}0
\def\ppnumber{\vbox{\baselineskip14pt
}}
\def\ppdate{MAD-TH-08-05, RUNHETC-2008-03}
\date{}

\author{
Gary Shiu$^1$,
Gonzalo Torroba$^2$, Bret Underwood$^1$\\
{\normalsize \it and}\\
Michael R. Douglas$^{2,3,\&}$\\[7mm]
{\normalsize $^1$Department of Physics, University of Wisconsin-Madison}\\
{\normalsize Madison, WI 53706, USA}\\
{\normalsize $^2$NHETC and Department of Physics and Astronomy}\\
{\normalsize Rutgers University. Piscataway, NJ 08854, USA}\\
{\normalsize $^3$Simons Center for Geometry and Physics,
 Stony Brook University}\\
{\normalsize $^\&$I.H.E.S., Le Bois-Marie, Bures-sur-Yvette, 91440 France}\\
[2mm]}

\title{\bf \LARGE Dynamics of Warped Flux Compactifications}

\vskip 1cm
\maketitle
\vskip 1cm

\begin{abstract} \normalsize
\begin{center}

\end{center}
\noindent We discuss
the four dimensional effective action for type IIB
flux compactifications, and obtain the quadratic terms 
taking warp effects into account. The analysis
includes both the 4-d zero modes and their KK excitations, which
become light at large warping. We identify an `axial' type gauge for
the supergravity fluctuations, which makes the four dimensional
degrees of freedom manifest.  The other key ingredient is the
existence of constraints coming from the ten dimensional equations of
motion. Applying these conditions leads to considerable
simplifications, enabling us to obtain the low energy lagrangian
explicitly. In particular, the warped K\"ahler potential for metric
moduli is computed and it is shown that there are no mixings with the
KK fluctuations
and the result differs from previous proposals. The four dimensional potential contains a
generalization of the Gukov-Vafa-Witten term, plus usual mass terms
for KK modes.
\end{abstract}

\bigskip
\newpage

\tableofcontents

\vskip 1cm

\section{Introduction}\label{sec:intro}

Understanding warp effects in string theory is important both for
theoretical issues (e.g. gauge/string dualities) and for making
more precise four dimensional predictions. 
The original motivation for warped geometries comes from considering
stacks of large numbers of branes, and reveals deep dualities between
supergravities and gauge theories, as in the AdS/CFT
correspondence~\cite{ads}.  A well studied duality with $\mathcal N=1$
supersymmetry involves the gauge theory of D5-branes wrapping an
isolated two-cycle of a Calabi-Yau compactification (the resolved
conifold, see for example \cite{dv}).
It is dual to a ten dimensional supergravity solution found
by Klebanov and Strassler~\cite{ks}.  In this example it is clear that
nontrivial variation of the warp factor is related to the nonzero beta
function of the theory, while the deformation of the conifold is
associated to the mass gap.  Some analysis of fluctuations around this
background has been made~\cite{italian, ksfluct1, ksfluct2, Gubser:2004qj}.

Warped geometries also play an important role in supersymmetry
breaking scenarios from string theory. For instance, placing
antibranes at the end of the conifold,~\cite{kpv} found a supergravity
dual of a nonsupersymmetric field theory. Other approaches include
brane-antibrane systems ~\cite{ag, dst}, and various other geometrical
effects~\cite{obsgeom, obsgeom2}. There have also been recent
developments in metastable vacua, following the work of~\cite{iss};
see, for example~\cite{meta1}-\cite{broek}. Overall, these works
suggest that strongly warped supergravities which break supersymmetry
are dual to dynamical supersymmetry breaking in gauge theories.

These works studied the local geometry near small cycles in the
compact manifold, as this is what is relevant for gauge-gravity
duality.  The global study of warped compactification is far more
difficult, as one cannot find exact solutions in this case.
Nevertheless one can make progress by combining a well chosen ten
dimensional ansatz with results obtained from some corresponding four
dimensional effective field theory.  In particular, one can look for
solutions obtained from a six dimensional Calabi-Yau manifold, by turning
on additional fields, and adding conformal and warp factors to the metric.
While explicit Calabi-Yau metrics are not known, a great deal of technology
has been developed to compute low energy observables anyways, which might
be adapted to these warped solutions.

A prototype here is the work of Giddings, Kachru, and Polchinski
(GKP) \cite{gkp}, who considered type IIB flux
backgrounds
satisfying certain BPS-type
conditions (see~\cite{Grana:2005jc, review1, review2, freyThesis} for
reviews and~\cite{becker, verlinde, sethi} for earlier work).
They were able to derive many properties of the
dimensionally reduced theory, including the flux superpotential and
the generation of large hierarchies. An important lesson from GKP is
that the dimensionally reduced action encodes many features of the
full theory in a simple way.

The actual results of GKP did not include the warp corrections to the
four dimensional effective action.  While this is consistent in the
limit of large volumes or small fluxes, clearly one would like to go
on and derive these corrections.  

To begin with some general comments, the warp factor is not
holomorphic and thus one expects it not to affect holomorphic
quantities such as the superpotential and gauge kinetic terms.\footnote{A caveat here is that the 4d superspace formulation of warped compactifications is not yet fully understood. It has been suggested~\cite{gid2} that some of the holomorphic K\"ahler coordinates should be expressed as periods of warped forms. See also~\cite{grimm-hitchin, 10to4}.}
However, in general it will affect non-holomorphic quantities such as
the K\"ahler potential.  Furthermore, since the warp factor is
produced by backreaction from fluxes and branes, one cannot hope
to find a general formula for the K\"ahler potential purely within the
moduli sector; rather it must couple moduli and other matter.  One
approach is to keep the warp factor as a six-dimensional field which
is determined by the fluxes and matter by solving a six-dimensional
equation, and then use it in defining the 4d effective K\"ahler
potential.  If this 6d equation uniquely determines the warp factor,
we can still regard the formalism as a 4d effective action.

A first attempt at a quantitative analysis was made in~\cite{gid1},
where a conjecture was made for the effect of warping on the complex
structure moduli space metric (see also \cite{FreyPolchinski} for an earlier
study of warped moduli space). Further study of the fluctuating modes
was made in \cite{gid2,alwis, Kahler, frey}.  However, these analyses became
extremely involved due to mixings between the various ten dimensional
modes and the need for ``compensator fields,'' and simple results were
not obtained.  For example, the validity of the conjecture of
\cite{gid1} remained unclear.  This was further studied in the language
of generalized complex geometry in \cite{10to4}.

Another fundamental difficulty is that, given strong warping, KK modes can
acquire masses of the same scale as the stabilized complex moduli, so
they may need to be included in an effective description. At present, a
consistent effective action is unknown, and there are doubts regarding
the validity of a four dimensional description~\cite{gid2, frey, burges}.

The aim of our work is to perform a consistent dimensional reduction
in the presence of warping, including both the 4d zero modes and the
first KK excitations. The main obstacles which will be faced are
finding a well-defined compactification procedure (we will point out
the subtleties in a moment), and identifying the correct 4d degrees
of freedom from the highly coupled 10d fluctuations. Before
starting, we review the warped supergravity backgrounds which will be
considered, and then summarize our work.

\subsection{Review of GKP ansatz}

Following \cite{gkp}, one starts with the bosonic action
$$
S_{IIB}=S_{EH}+S_{matter}=\frac{1}{2\kappa_{10}^2} \int
d^{10}x\,\sqrt{-g}\, \Big\{R-\frac{\partial_M \tau \partial^M \bar
\tau}{2 (\rm Im \tau)^2} -\frac{G_3 \cdot \bar G_3}{12 {\rm Im
\tau}} -\frac{\tilde F_5^2}{480}\Big\}+
$$
\begin{equation}\label{eq:Sb1}
-\frac{i}{8\kappa_{10}^2} \int \frac{C_4 \wedge G_3 \wedge \bar
G_3}{{\rm Im \tau}}+S_{loc} .
\end{equation}

We take space-time to be a warped product of $\RR^{3,1}$ with
a six dimensional K\"ahler manifold $M$.  Let
$\tilde g_{mn}$ be a Ricci-flat K\"ahler metric on $M$, then 
the metric ansatz is
\begin{equation}\label{eq:wmetric1}
ds_{10}^2=e^{2A(y)}\eta_{\mu \nu} dx^\mu dx^\nu+e^{-2A(y)} \tilde
g_{mn} dy^m dy^n\,.
\end{equation}
The field strengths are chosen to preserve Lorentz invariance and
self-duality is imposed on the five form,
\begin{equation}\label{eq:g3}
G_3=\frac{1}{6} G_{mnp}(y) dy^m dy^n dy^p=F_3-\tau H_3\,,
\end{equation}
\begin{equation}\label{eq:f5}
\tilde F_5= \partial_m \alpha(y) (1+\star) dy^m dx^0 dx^1 dx^2 dx^3\,.
\end{equation}
Since we are interested in flux compactifications, we
will assume that $F_3$ and $H_3$ are three-forms in nontrivial classes of
$H^3(M, \mathbb Z)$.

The warp factor is of the general form
\begin{equation}\label{eq:c}
e^{-4A(y)}=c+e^{-4A_0(y)}
\end{equation}
where the dimensionless parameter $c$ is related to the total volume 
by $V_{CY} \sim \alpha'^3\,c^{3/2}$, and $A_0$ is produced by 
matter sources (fluxes or branes).  The large volume limit $c \gg
e^{-4 A_0}$ corresponds to a small number of fluxes or branes, where
backreaction may be ignored.  In the present work we will address the
general case, which includes the strongly warped limit $c \ll
e^{-4A_0}$. Notice also that in this regime the $c$ fluctuation does not coincide with the 
usual parameterization of the universal K\"ahler modulus as an overall scaling
of the underlying CY~\cite{gid2}.

Furthermore, we will concentrate on the BPS backgrounds of~\cite{gkp}, 
which are analogous to the brane metrics of AdS/CFT.  These satisfy
\begin{equation}\label{eq:bps}
\alpha=e^{4A}\;,\;\,\frac{1}{4} (T^m_m-T^\mu_\mu)^{loc}=T_3
\rho_3^{loc}\,.
\end{equation}
As a result, the equations of motion are verified automatically if
the three form flux is imaginary self-dual (ISD):
\begin{equation}\label{eq:isd}
\star_6 G_3=i G_3\,.
\end{equation}
Similar conditions may be written for backgrounds sourced by
anti-branes, with the flux becoming IASD.  As pointed out by
GKP, the ISD condition fixes the complex structure moduli and leads to 
a constant background dilaton.

Small fluctuations around this background should be described by a 4d
$\mathcal N=1$ supergravity effective action.  A reasonable starting
point for this is to take the $\mathcal N=2$ supergravity obtained by
KK reduction of IIb supergravity on $M$, and then apply an orientifold
projection.
This is expected to be a good description in the large volume limit, 
with small quantum corrections.  Furthermore, we are assuming that
backreaction can be ignored and thus the warp factor is essentially
constant.  This will be true if we
take the limit $\alpha' \to 0$ while holding the number of flux units $N$
fixed. Conversely, large $N$ dualities or large hierarchies arise, even
in the $\alpha' \to 0$ limit, when $\alpha' N$ is held fixed.

In this limit, the 4d K\"ahler potential for
the metric moduli takes the well-known form,
\begin{equation}\label{eq:unwarped-K}
K=-3 \log (-i(\rho+\bar{\rho})) - \log \big(-i \int J\wedge J\wedge J\big)
	-\log \big(-i \int_M \Omega\wedge \bar{\Omega}\big)\, .
\end{equation}
Fluxes generate a scalar potential for the complex moduli,
\begin{equation}
{\mathcal V} = -\frac{1}{2\kappa_{10}^2\mbox{Im}\tau} \int_M G_3\wedge (\star_6\bar{G}_3+i \bar{G}_3)
\end{equation}
which vanishes in the ISD case.
From the scalar potential and K\"ahler potential we can infer that the superpotential 
for the complex moduli is of the Gukov-Vafa-Witten type \cite{gkp, gvw, becker, tv},
\begin{equation}
W_{GVW} = \int_M \Omega\wedge G_3
\label{eq:GVWSuperpotential}
\end{equation}
while the K\"ahler moduli
only receive non-perturbative superpotential contributions~\cite{kklt}.

\subsection{Towards an effective description of warping}

Steps towards including the effects of warping were taken in~\cite{gid1}.
The simplest is to change the relation between the 10d Planck
scale, the 4d Planck scale, and the volume of the internal manifold,
to
$$
\frac{M_{Pl,4}^2}{M_{Pl,10}^2} = 
\left(\frac{M_{Pl,10}}{2\pi}\right)^6 V_W ,
$$
where $V_W$ is the ``warped volume'' of the internal manifold,
\begin{equation}\label{eq:Vw}
V_W=\int d^6y \sqrt{\tilde g_6}\,e^{-4A}\,.
\end{equation}

Then, the potential and kinetic terms of the 4d effective action
were computed by KK reduction.  In particular, the kinetic terms for
the moduli were obtained by varying the Einstein-Hilbert action
around the metric ansatz \eq{wmetric1}, obtaining
\begin{equation}
G_{\alpha\bar{\beta}}^{W} = \partial_\alpha \partial_{\bar{\beta}} K_{W} = \int d^6y \sqrt{\tilde{g}_6}\ e^{-4A} 
\partial_\alpha \tilde{g}_{mn}\, \partial_{\bar{\beta}} \tilde{g}^{mn} \,.
\label{eq:DGMetric}
\end{equation}
Then, substituting in the variations of the 6d metric with the moduli,
it was found that the warp factor corrections change
\eq{unwarped-K} to 
\begin{equation} \label{eq:DGKahler}
K_{W} \stackrel{?}{=} -3 \log (-i(\rho+\bar{\rho})) - \log \big(-i \int_M e^{-4A} J\wedge J\wedge J\big)
- \log \big(-i \int_M e^{-4A} \Omega\wedge \bar{\Omega}\big)
\end{equation}

However, in \cite{gid2} and other works, it was realized that this
analysis was somewhat oversimplified.  One reason for this was that,
in generalizing the KK ansatz to fields which vary in the four dimensions,
one often needs to add terms which depend on derivatives of the fields.
These ``compensators'' were discussed at some length in \cite{gid2}, with
the conclusion that they are necessary, and do contribute to the kinetic
terms, placing \eq{DGKahler} in some doubt.

As mentioned before, KK modes becoming light in regions of strong
warping also need to be included in the effective action.  While there
are a few compactifications in which a consistent truncation to a
small subset of modes is possible \cite{pope}, usually this is not the
case, and there is no reason to believe it is so for an arbitrary
Calabi-Yau compactification.  Furthermore, in \cite{gid2} it was
claimed that mixings between the zero modes and their KK fluctuations
are expected even at the level of the kinetic terms. Unlike the usual
Calabi-Yau case, this would imply that the low energy dynamics
truncated to metric moduli is inconsistent!

\subsection{Some applications of warped effective theories}

Warped compactifications are ubiquitous in the constructions of
phenomenologically attractive string vacua, both for applications to
particle physics and to cosmology.
Thus, understanding effective theories in the presence of warping is essential
for investigating the physics of such
vacua where a hierarchy of scales can be generated.
For example,
in the construction of de-Sitter vacua
in \cite{kklt},  the existence of a strongly warped region is a key ingredient
for the uplifting procedure. To properly describe the physics around these stabilized vacua,
it is important to understand
corrections to the effective theory due to warping.  Furthermore, many
crucial
aspects of
supersymmetry breaking in warped backgrounds -- including its mediation
mechanisms \cite{SmallNumbers,Mirage}, sequestering of supersymmetry
breaking \cite{Sequestering}, and the computations of soft terms
-- depend on
quantitative
details of
the warped effective theory.
In fact, if warping is responsible for generating the electroweak hierarchy,
a plausibly distinct signature
of warped throats  is the production of massive KK resonances at colliders;
the precise signatures can depend on details of the warped
geometry and of the interactions involving the KK modes \cite{WarpedLHC}.

On the other hand,
explicit models of string inflation often
make use of D-branes in warped throats since the
warp factor can help in flattening the inflaton potential \cite{KKLMMT}.
Predictions of these
models depend strongly on the underlying
effective action describing the closed string moduli as well as the
moduli
parametrizing the position of the mobile
D-brane  \cite{DbraneInflation}. Hence, modifications to the effective
action due to warping could have interesting phenomenological and
model building consequenses.  Furthermore, the end of D-brane inflation typically
results in a brane and antibrane annihilating, with most of the energy
ending up in KK modes of the warped throat \cite{KofmanYi,Chialva}.
Reheating of the universe occurs when the KK modes decay on the branes
which realize the Standard Model, and thus the details of reheating depend on
the properties of the KK modes
\cite{KofmanYi,Chialva,WarpedReheating,tye,Kofman2}.
Therefore, the mixing among KK
modes can have
important consequences for these issues.

In light of these applications, it is an important problem to compute
the dimensionally reduced effective action for warped
compactifications, taking the previously identified challenges into
account.

\subsection{Summary of the paper}

In this work we develop the theory of warped compactification as
follows. In section \ref{sec:fluct} the theory is analyzed from a ten
dimensional perspective in the axial gauge $g_{\mu m}=0$. We identify
the 10d fluctuations and compute their equations of motion. In
subsection \ref{subsec:10dpersp} we define, for each type of
fluctuation, a basis of ``warped'' internal wavefunctions which will
be used throughout the work.  An important result is that metric
fluctuations must satisfy the constraints
Eqs.~(\ref{eq:PhysicalGauge1}) and (\ref{eq:PhysicalGauge2}), which
follow from the $(\mu \nu)$ and $(\mu m)$ Einstein equations. We
provide a general formula for the effective action in
Eq.(\ref{eq:Seff1}).

Section \ref{sec:kinetic} presents the four dimensional kinetic terms
for the different fluctuations. It is argued that there are no terms
with two space-time derivatives mixing the metric moduli with any of
the other light modes, in the basis of ``warped'' KK modes defined in
subsection \ref{subsec:10dpersp}. Therefore, even in the strongly
warped limit it is consistent to study the propagators associated to
such moduli independently of the other fields.  In Eq.~(\ref{eq:Gs})
we present the warped K\"ahler potential for complex moduli and
address various puzzles related to this term.

In section \ref{sec:masses} we study the geometrical KK masses,
nonvanishing even for zero fluxes. Subsection \ref{subsec:mass-scalar}
provides a mathematical and physical justification for the use of the
previous basis of ``warped'' eigenmodes. We show that these mass terms
do not mix the metric moduli with KK excitations, while we do find
quadratic couplings between massive graviton and internal metric
modes. This suggests that a certain geometrical Higgs mechanism may be
at work in the warped compactifications being analyzed.

Finally, in section \ref{sec:fluxmass} we find the flux potential for
the metric fluctuations, including KK modes. The result, given in
Eq.~(\ref{eq:Sv2}), is a warped generalization of the
Gukov-Vafa-Witten potential. Interestingly, this potential exhibits
possible mixings between the moduli and their KK tower. Our analysis
applies to nontrivial axio-dilaton fluctuations as well, which also
mix with the metric moduli. In the appendices we collect various
useful formulas and show some of the computations in detail.

\section{From 10 to 4: Warped Fluctuations and Effective Action}
\label{sec:fluct}

In this section, the general procedure for obtaining the effective
action for warped flux compactifications will be described.  First, we
identify the ten dimensional fluctuations of the metric and matter
which give rise to 4d fields. Key to our analysis is the choice of a
supergravity gauge fixing condition, which simplifies the problem
considerably. Examining the equations of motion, we find the existence
of {\it constraint equations}; these relate different fluctuations to
each other and fix residual gauge transformations. Finally, we provide
a general formula for computing the effective action.

\subsection{10d perspective for fluctuations in IIB \mbox{supergravity}}\label{subsec:10dpersp}

From a ten-dimensional point of view, the dynamics follows by
considering infinitesimal fluctuations around the previous backgrounds
in which the moduli are spacetime dependent, and then solving the
corresponding equations of motion.  The zero mode sector includes the
complex and K\"ahler moduli $u^I=(\rho^i, S^\alpha)$, the 4d graviton
$h_{\mu \nu}(x)$, the axio-dilaton $\tau_0(x)$ (both are constant on
the internal manifold), and the various massless $p$-form fields
coming from decomposing $(C_2, B_2, C_4)$ into harmonic forms. For
each of them we will include the corresponding tower of KK
excitations.

We will take the fluctuations of the 10-dimensional metric, in the presence of 
dynamical moduli, to have the form 
\begin{equation}
\delta (ds^2) = \delta g_{\mu\nu} dx^\mu dx^\nu + \delta g_{mn} dy^m dy^n\, .
\end{equation}
where
\begin{equation}\label{eq:metricfluct1a}
\delta g_{\mu \nu}(x,y)=e^{2A(y)}[2 \,\delta A(x,y) \eta_{\mu
\nu}+\delta_K g_{\mu \nu}(x,y)]\,,
\end{equation}
\begin{equation}\label{eq:metricfluct1b}
\delta g_{mn}(x,y)=e^{-2A(y)}[-2 \,\delta A(x,y) \tilde
g_{mn}(y)+\delta \tilde g_{mn}(x,y)]\,.
\end{equation}
Here $\delta_K g_{\mu \nu}$ are 4d graviton KK modes (which are not
necessarily transverse-traceless), while $\delta \tilde g_{mn}$ encode
the metric moduli $u^I$ and their KK modes.  Since the warp factor
depends on the moduli, a fluctuation $\delta \tilde g_{mn}$ induces in
turn a variation $\delta A=u^I \partial A/\partial u^I$. It is not
consistent to set $\delta A=0$.

In general there are also off-diagonal moduli-dependent fluctuations
of the form $\delta g_{\mu m} \sim \partial_\mu u(x)
B_m(y)$~\cite{gid2, gray}. The reason for this particular form is that
in the limit in which the moduli become constant, we should recover
the background (\ref{eq:wmetric1}).  These are gauge dependent and
as we will discuss elsewhere \cite{DT} we can remove them by going to
an ``axial'' type gauge
\begin{equation}\label{eq:physgauge}
 \delta g_{\mu m}=0\,.
\end{equation}
This choice will make the four dimensional degrees of freedom
manifest. As explained in the next subsection, the remaining
diffeomorphisms that preserve this condition are fixed by the
equations of motion.

To isolate the zero modes from their KK partners, we expand the metric fields
in a basis of eigenmodes for the internal manifold:
\begin{equation}\label{eq:gravitonKKExpansion}
\delta_K g_{\mu \nu}(x,y)=\sum_{I_1}\,h^{I_1}_{\mu \nu}(x)
Y^{I_1}(y)\, .
\end{equation}
\begin{equation}\label{eq:internalKKExpansion}
\delta \tilde g_{mn}(x,y)=\sum_{I_2}\,u^{I_2}(x)\,Y_{mn}^{I_2}(y)\,,
\end{equation}
The multi-index $I_i$, $i=1,2$, runs over the different types of metric
fluctuations and, for each type of fluctuation, over the 4d KK tower. $I_i=0$
gives the zero mode of the appropriate Laplacian on the internal manifold.
To be precise, in \eq{internalKKExpansion} we must specify a gauge.  This
will be \eq{PhysicalGauge2} as we discuss below.

We certainly have the freedom to choose different complete bases of
internal wavefunctions; this amounts to making field redefinitions in
the four-dimensional theory. Different choices represent the extra
dimensions in rather different ways and our aim is to find the one
that yields the simplest description. For instance, for a scalar mode
in six dimensions, there are two natural possibilities, either
\begin{equation}\label{eq:eigenvector}
\tilde \nabla^2\,Y_i(y)=e^{-4A(y)}\,\lambda_i^2\,Y_i(y)\,,
\end{equation}
or the usual unwarped KK modes,
\begin{equation}\label{eq:unweigenvector}
\tilde \nabla^2\,\mathcal Y_i(y)=\nu_i^2\,\mathcal Y_i(y)\, .
\end{equation}
However, since in a warped compactification the mass
eigenmodes will be given by $Y_i(y)$, it is much simpler to
expand the ten dimensional fields in the first basis of ``warped'' KK modes.

One can check that \eq{eigenvector} 
is a well-defined Sturm-Liouville problem; thus if we use the inner
product in which this linear operator is self-adjoint, 
non-degenerate eigenvectors will be orthogonal and we can orthogonalize
degenerate eigenvectors.  Thus we can choose a basis in which
\begin{equation}
\int d^6y \sqrt{\tilde g_6}\,e^{-4A}\,Y_i(y)
Y_j(y)=G\,\delta_{ij}\,.
\end{equation}
In the following, this is made explicit in each sector.

\vskip 1mm

\noindent{\bf Metric fluctuations}

We will take the eigenmodes  
(\ref{eq:gravitonKKExpansion}), (\ref{eq:internalKKExpansion})
to be solutions to the respective eigenvalue equations,
\begin{equation}
\label{eq:GravitonEigenmode}
\tilde \nabla^2\,Y^{I_1}=e^{-4A(y)}\lambda^2_{I_1}\,Y^{I_1}\,,
\end{equation}
\begin{equation}
\label{eq:InternalEigenmode}
\frac{1}{2}\tilde \Delta_L\,Y^{I_2}_{mn}(y)=\delta \tilde
G_{mn}=e^{-4A(y)}\lambda_{I_2}^2\,Y^{I_2}_{mn}(y)\,;
\end{equation}
here $\tilde \Delta_L$ is the Lichnerowicz laplacian for $\tilde
g_{mn}$. The eigenmode expansions (\ref{eq:InternalEigenmode}, \ref{eq:GravitonEigenmode}) lead
to orthogonality relations between different modes
\begin{equation}\label{eq:Ggraviton}
\frac{1}{2}\int d^6y \sqrt{\tilde{g}_6} e^{-4A} Y^{I_1}\, Y^{J_1} = {\mathcal M}^{kk} \delta_{I_1,J_1}\, .
\end{equation}
\begin{equation}\label{eq:orthon}
\int d^6 y \sqrt{\tilde g_6} \,e^{-4A(y)}\,Y^{I_2}_{mn}(y) \bar
Y^{J_2\,\widetilde{mn}}(y)=G^{(u)}\,\delta_{I_2,J_2}\,.
\end{equation}
Indices with tildes are raised with $\tilde g^{mn}$.
\vskip 1mm

\noindent{\bf Dilaton fluctuations}

Fluctuations of the dilaton are of the form $\tau = \tau_0 + \delta\tau(x,y)$, 
where ${\rm Im}\,\tau_0=g_s^{-1}\gg 1$ and is constant in the internal space.
Since the axio-dilaton is a scalar from the six-dimensional point of view, its expansion in KK modes
is the same as for the graviton:
\begin{equation}\label{eq:tauexp1}
\delta\tau = \delta\tau_0(x)+ \delta_K \tau(x,y) = \delta\tau_0(x) +\sum_{I_1}\,t^{I_1}(x)Y^{I_1}(y)\, .
\end{equation}
The orthogonality relation for the dilaton KK modes then reads,
\begin{equation}\label{eq:orthog}
-\frac{1}{4(\Imtau_0)^2}\int d^6y \sqrt{\tilde g_6}\,e^{-4A}\,Y^{I_1}(y)
Y^{J_1}(y)\equiv {\mathcal M}^{kk}_\tau\,\delta_{I_1 J_1}\,.
\end{equation}

\vskip 1mm
\noindent{\bf $p$-form fluctuations}

The various antisymmetric tensors have an expansion of the form
\begin{eqnarray}\label{eq:pKK}
&&F_3 \to F_3+ d\,\delta C_2\;,\;H_3 \to H_3+ d\,\delta B_2\;,\; \nonumber \\
&&F_5 \to F_5+ d\,(\delta\alpha)\wedge d^4x+d\,\delta C_4\,.
\end{eqnarray}
$(F_3, H_3, F_5)$ denote the background GKP values, while the derivative variations include the zero modes 
(harmonic forms) and their KK excitations.  Note that because of the background BPS
relation $\alpha = e^{4A}$, fluctuations of the 4-form are induced by fluctuations of the warp factor (which
are in turn induced by fluctuations of moduli), e.g. $\delta \alpha = \delta e^{4A}$.

As with the metric, when the moduli are promoted to spacetime dependent fields in general one should include terms in the
$p$-form fluctuations proportional to spacetime derivatives of the moduli,
$\delta C_2\sim du^I \wedge T_I$, $\delta C_4 \sim du^I \wedge S_I^{(3)}+..$.  As shown in Appendix \ref{app:flux}, solving
the flux equations of motion leads to the identification of these fluctuations as pure gauge, and we will fix the gauge freedom by choosing them to vanish. 
This is the analog of Eq. (\ref{eq:physgauge}). This is a crucial simplification and is particular to GKP backgrounds. In more general settings~\cite{gid2}, the previous is not true and the ten dimensional analysis becomes quite involved.

In principle, a nontrivial warp factor could induce mixings between the four dimensional scalars coming from these $p$-forms and the metric moduli. However, we will argue in section \ref{subsec:nonmixp} that this is not the case; 
for this reason, we will not perform a full analysis of this sector.

\subsection{Ten dimensional fluctuated equations}\label{subsec:comps}

The linearized equations for the fluctuations are
\begin{equation}
\delta G_{MN}=\kappa_{10}^2 \delta T_{MN}\;,\;d \delta
F_5=\delta(H_3\wedge F_3)+2\kappa_{10}^2 \delta (T_3
\rho_3^{loc})\,,
\label{eq:10dEOM}
\end{equation}
\begin{equation}
d\, \delta\big[e^{4A}(\star_6 G_3-iG_3) \big]=0\,, \ \delta (\star_{10} F_5) = \delta F_5\, ,
\label{eq:flux10deom}
\end{equation}
\begin{equation}\label{eq:flucteom}
\nabla_M\nabla^M \delta_K\tau = - \frac{i}{12} \delta (G_3\cdot G_3)\, ,
\end{equation}
using the usual formula
\begin{equation}\label{eq:fluctR}
\delta R_{MN}=-\frac{1}{2} \nabla^P \nabla_P \delta
g_{MN}-\frac{1}{2} \nabla_M \nabla_N g^{PQ}\delta g_{PQ}+\frac{1}{2}
\nabla^P \nabla_M \delta g_{NP}+\frac{1}{2} \nabla^P \nabla_N \delta
g_{MP}\,.
\end{equation}
These equations are discussed in more detail in the Appendix.

An important point of this work is to recognize that some of these equations contain at most first order space-time derivatives. These ``constraints'' are familiar from systems with gauge redundancies. For example, in general relativity, $G_{0 i}=T_{0 i}$ does not contain second order time derivatives, and so 
it has to be satisfied by any consistent solution, at all times~\cite{wald}. Another famous example 
of this type is Gauss' law in electrodynamics. We now identify such initial value constraints for warped supergravities. They will play a crucial role in simplifying the four dimensional action.

One such constraint comes from the $(\mu \nu)$ Einstein equation\footnote{A similar expression is found in the
dimensional reduction of the radion in the Randall-Sundrum model \cite{RadionDimReduct}.},
\begin{equation}
\nonumber
\delta G_{\mu\nu} - \kappa_{10}^2\delta T_{\mu\nu} = 
\partial_\mu \partial_\nu (4 \delta A - \frac{1}{2}\delta \tilde{g}) + \eta_{\mu\nu} [...] = 0\, .
\end{equation}
Subtracting out the trace of this equation, we obtain the constraint
\begin{equation}
\delta A = \frac{1}{8}\delta\tilde{g}\,.
\label{eq:PhysicalGauge1}
\end{equation}
Computationally, this is a very useful relation, allowing us to replace the potentially complicated fluctuation $\delta A$ by $\delta \tilde g$.
This equation has a nice physical interpretation as the invariance of the warped volume Eq.~(\ref{eq:Vw}) with respect to fluctuations of the moduli, $\delta V_W = 0$.

Another constraint comes from the $(\mu m)$ Einstein equation
\begin{equation}
\nonumber
\delta G_{\mu m} - \kappa_{10}^2 \delta T_{\mu m} = 
\frac{1}{2}\partial_\mu \left[\tilde{\nabla}^n (\delta\tilde{g}_{mn} - \frac{1}{2} \tilde{g}_{mn} \delta\tilde{g})
-4\partial^{\tilde n} A\,\delta \tilde g_{mn}\right] = 0\, ,
\end{equation}
which indicates that the condition 
\begin{equation}
\tilde \nabla^n\big(\delta \tilde g_{mn}-\frac{1}{2}\tilde
g_{mn}\,\delta \tilde g\big)=4 (\partial^{\tilde n}A)\,\delta \tilde
g_{mn}\,
\label{eq:PhysicalGauge2}
\end{equation}
must be satisified. This resembles a warped generalization of the
harmonic gauge and, indeed, it fixes the remaining six dimensional
diffeomorphisms that leave the axial gauge (\ref{eq:physgauge})
invariant. However, it must be stressed that this is not a choice, but
a constraint imposed by the dynamics. Fluctuations that do not satisfy
this consistency expression will not give a consistent ten dimensional
solution.

Finally, another set of constraints follows from the moduli-dependent
parts of the five form and three form equations of motion, self
duality, and Bianchi identities,
(\ref{eq:10dEOM}-\ref{eq:flux10deom}).  We will not write these out in
detail here -- explicit use of them will be made in the Appendix.

\subsection{Four dimensional effective action}
\label{subsec:fluctaction}

As the final step in our general discussion, we wish to understand the
dynamics from a four dimensional action principle, by first
compactifying the supergravity action on a background satisfying the
ten-dimensional equations of motion, and then integrating over the
internal coordinates, along the lines of~\cite{candelas}.

There are some known subtleties in doing this.  First, the type IIB
supergravity action is ill-defined due to the self-duality of the five
form. The procedure which will be followed here is to project out half
of the 4-dimensional degrees of freedom of the five form and double
the coefficients of the $F_5^2$ and Chern-Simons term \cite{gid1}.
Second, the Gibbons-Hawking-York term \cite{ghy} must be included in
the dimensional reduction to cancel certain total derivative terms of
the variation of the gravitational action.

The dimensionally reduced effective action is then obtained by expanding 
\begin{equation}\label{eq:generalS}
S=\int_M d^{10} x \sqrt
g\,(g^{MN}\,R_{MN}+\mc L_{matter})
+2\int_{\partial M}\,d^9x\,\sqrt h\,K\,
\end{equation}
to second order in fluctuations. After some algebra, the result is,
\begin{equation}\label{eq:Seff1}
S_{eff}=\frac{1}{4\kappa_{10}^2} \int d^{10} x\,\sqrt g\,\Big[ - (\delta g)^{MN}
\big(\delta G_{MN}-\delta T_{MN} \big)+\delta^2 \mc L_{matter}\Big]+\mc O (\delta g^3)\,,
\end{equation}
where $\delta^2 \mc L_{matter}$ represents second order fluctuations of the flux and dilaton terms in the effective
action with respect to the flux and dilaton fields.
This reproduces the correct ten dimensional equations of motion. 
To perform the dimensional reduction, one expands the fluctuations in
terms of internal eigenmodes and integrates over the compactification
space, while imposing the constraint equations
(\ref{eq:PhysicalGauge1}, \ref{eq:PhysicalGauge2}) derived in the
previous subsection.

Formally, in this process we are including all the (infinite tower of)
KK modes, to quadratic order.  The details of each particular region
of strong warping will then determine a truncation of the KK tower to
keep only the lightest modes. Higher order interactions may also be
analyzed by further expanding, for instance, the Einstein-Hilbert
term. The existence of trilinear couplings between moduli and KK modes
may have interesting consequences. In the rest of the paper we will
study this effective action, sector by sector, in detail. We present
the full computation in Appendix \ref{app:full}.

\section{Warped Kinetic Terms and Moduli Space Metrics}\label{sec:kinetic}

This section is devoted to the analysis of the terms in the effective action that contain space-time derivatives. The main result will be that the kinetic terms for all sectors decouple and are diagonal in the KK mode expansion,
$$
\mc L_{kin} = G^{(u)}\sum_{I_2}  u^{I_2}\,\Box \bar{u}^{I_2} + \mc M^{kk}_{h} \sum_{I_1}   E^{\mu\nu}_{I_1}\, h^{I_1}_{\mu\nu} + \mc M^{kk}_\tau \sum_{I_1}   t^{I_1}\,\Box \bar t^{I_1}\,.
$$
Here, $E_{\mu \nu}$ is the linearized four dimensional Einstein tensor and $I_i$ runs over the different moduli and their KK towers. 
This implies that (even with a nontrivial warp factor) the propagators for the moduli and for the light KK modes do not mix.
The diagonality of the kinetic terms implies that the K\"ahler potential 
is also diagonal to quadratic order in the fluctuations of the moduli and the dilaton (and their KK modes),
\begin{equation}
K = G^{(u)}\sum_{I_2}  u^{I_2} \bar{u}^{I_2} + \mc M^{kk}_\tau \sum_{I_1}   t^{I_1}  \bar t^{I_1}\,.
\end{equation}

\subsection{Axio-dilaton and $p$-form modes}\label{subsec:Ktau}

As a warm-up, we first consider the kinetic terms for the axio-dilaton
and the $p$-forms.  Here the dimensional reduction is simple but
nonetheless ilustrates some of the features that will appear in the
more involved analysis of metric fluctuations.

From Eq.~(\ref{eq:Sb1}), the axio-dilaton is a ten-dimensional scalar field with a nonlinear metric. Expanding around the background value $\tau_0$ according to (\ref{eq:tauexp1}), the kinetic term turns out to be
\begin{equation}\label{eq:Sefftau}
\mc L_{kin}^{(\tau)}= \sum_{I_1,J_1} G^{(\tau)}_{I_1 J_1} t^{I_1}(x) \Box \bar t^{J_1}(x)
\end{equation}
where the warped metric is
\begin{equation}\label{eq:Gtau}
G^{(\tau)}_{I_1 J_1}=\frac{1}{4({\rm Im}\, \tau_0)^2\,V_W}\,\int d^6y \sqrt{\tilde g_6}\,e^{-4A(y)} Y_{I_1}(y) Y_{J_1}(y)\, .
\end{equation}
Notice that the dilaton metric (\ref{eq:Gtau}) is proportional to the orthogonality relation on the internal
space (\ref{eq:orthog}), so that the K\"ahler potential, to quadratic order, becomes (taking
$\mc M^{kk}_\tau \equiv G^{(\tau)}_{I_1 I_1}$)
\begin{equation}
K^{(\tau)} = \mc M^{kk}_\tau\,\sum_{I_1}  t^{I_1}(x) \bar{t}^{I_1}(x)\, .
\label{eq:tauKahler}
\end{equation}
The conclusion is that, at this order, there is no mixing between the various dilaton KK modes in the kinetic term. Had we used the unwarped KK expansion of Eq.~(\ref{eq:unweigenvector}), the kinetic term would have exhibited complicated mixings.

\vskip 1.5mm

Next we discuss a rather different behavior, arising from the one-form KK modes of $C_4$:
$$
\delta C_4= \sum_I V^I_\mu(x) dx^\mu \wedge \chi^I(y)\,,
$$
where $\chi^{I}(y)=\frac{1}{3!}\chi^{I}_{mnp}\,dy^{mnp}$. This case is relevant for D-term supersymmetry breaking~\cite{dterm}. Replacing this in the $\tilde F_5^2$ term of the action (\ref{eq:Sb1}), we find
\begin{equation}\label{eq:SeffV}
\mc L_{kin}^{(V)}=G^{(V)}_{I J}\,F^{I}\wedge \star_4 F^{J}\,.
\end{equation}
The metric appearing here is 
\begin{equation}\label{eq:GV}
G^{(V)}_{I J}=\sum_{I,J}\frac{1}{4\,V_W}\,\int_M \chi_{I}(y) \wedge \star_6 \chi_{J}(y)\,,
\end{equation}
Therefore, the field space metric coincides with the unwarped one, with the result that the corresponding D-term cannot be made parametrically small by the large hierarchy of the throat. 
The field space metric $G^{(V)}$ for the massless mode can be shown to coincide with 
$\rm Im\,\partial^2 \mathcal F$, where $\mathcal F$ is the Calabi-Yau prepotential.

\subsection{Graviton fluctuations}\label{subsec:confK}

After having gained some intuition with the previous simpler sectors, we will now consider the metric fluctuations,
\begin{equation}\label{eq:metricfluct2}
\delta(ds^2)=e^{2A}\big[2 \delta A\, \eta_{\mu \nu}+\delta_K g_{\mu \nu} \big]dx^\mu dx^\nu+e^{-2A}\big[-2 \delta A \,\tilde g_{mn}+\delta \tilde g_{mn} \big] dy^m dy^n\,.
\end{equation}
The kinetic term for this sector follows from dimensionally reducing the Einstein-Hilbert part of the supergravity action, according to the prescription Eq.~(\ref{eq:Seff1}). 

The variation of the warp factor makes this computation highly nontrivial in at least two aspects. 
First, from the space-time variation we expect mixings between the trace part of the 
graviton mode\footnote{Note that we have not chosen the standard transverse traceless gauge for the graviton, 
which is in general not consistent with axial gauge \cite{gid2}.} and the moduli through $\delta A$, $\delta\tilde{g}$. 
Further, the internal metric 
variation is no longer proportional to $\delta \tilde g_{mn}$ (since it includes $\delta A$ fluctuations)
so the relation between a complex modulus of the underlying Calabi-Yau ($\delta \tilde G_{mn}=0$) and a 
zero mode of the full warped metric ($\delta G_{MN}=0$) may be very involved.

Our first result will be to show that, in spite of the possible couplings suggested by (\ref{eq:metricfluct2}), there are no space-time derivative mixings between $\delta_K g$ and $\delta \tilde g_{mn}$. The simplest way of understanding this is by doing a conformal transformation, and for this it is actually better to work with the metric containing the fluctuations to all orders:
\begin{equation}\label{eq:metricfluct4}
ds^2=e^{2A(x,y)}\,g_{\mu \nu}(x,y)\, dx^\mu dx^\nu+ e^{-2A(x,y)}\,\tilde g_{mn}(x,y)\,dy^m dy^n\,.
\end{equation}
The $x$ dependence in $A$ and $\tilde g_{mn}$ comes from promoting the moduli to space-time dependent fields and from their KK modes. The graviton is associated to the fluctuating metric $g_{\mu \nu}(x,y)$ and does not induce warp factor fluctuations. 

Consider the conformal transformation
\begin{equation}\label{eq:confrescaling}
ds^2=e^{2A(x,y)}\,d\hat s^2\;,\;\hat g_{\mu \nu}=g_{\mu \nu}\;,\;\hat g_{mn}=e^{-4A}\,\tilde g_{mn}\,,
\end{equation}
which leads to a change in the Ricci scalar ~\cite{wald}
\begin{equation}\label{eq:confrescaling2}
R\to e^{-2A}\big(\hat R+9 \times 8\, \hat g^{LM} \partial_L A\,\partial_M A\big)\,,
\end{equation}
after an integration by parts.
The conformally rescaled metric $\hat g_{MN}$ does not mix the KK graviton and warp factor, making the decoupling of these fluctuations manifest.
Dimensionally reducing $\hat R$ on this ansatz for $\hat g_{MN}$ leads to kinetic terms for the graviton and 
internal metric fluctuations without off-diagonal mixings.  Further, the spacetime derivative contribution of the extra term in
(\ref{eq:confrescaling2}) does not contain graviton pieces (at quadratic order). This proves 
that there are no space-time derivative mixings. This result is rederived in Appendix \ref{app:full}, by performing the computation in the original unrescaled metric.

After having established this, it is straightforward to compute the kinetic term for the graviton modes, since we only need to consider a metric perturbation
\begin{equation}
\delta (ds^2)=e^{2A(y)}\,\delta_K g_{\mu \nu}(x,y) dx^\mu dx^\nu\,.
\end{equation}
The result is a warped version of the linearized Einstein-Hilbert action around a flat background~\cite{wald},
\begin{eqnarray}\label{eq:kingrav}
S_{kin}^{(h)}&=&\frac{1}{2\kappa_4^2 V_W}\int d^4x \int d^6y \sqrt{\tilde g_6}e^{-4A} \delta_K g^{\mu \nu}\,\delta_K G^{(4)}_{\mu\nu}\nonumber\\
&=&\frac{1}{2\kappa_4^2 V_W}\int d^4x \int d^6y \sqrt{\tilde g_6}e^{-4A} \Big[ -\frac{1}{2}(\delta_K g^{\mu \nu}\,\Box \delta_K g_{\mu \nu}-\delta_K g \,\Box \delta_K g)+\nonumber\\
&+&\delta_K g^{\mu\nu}(\partial^\sigma \partial_{(\mu}\delta_K g_{\nu)\sigma}-\partial_\mu \partial_\nu\delta_K g)\Big]\,.
\end{eqnarray}
Indices are raised with $\eta^{\mu\nu}$, and $\delta_K g:=\eta^{\mu \nu}\delta_K g_{\mu\nu}$.
In harmonic gauge the last line vanishes and we recover the usual kinetic term; however, at this point we prefer to keep the graviton 
gauge 
arbitrary.

Expanding $\delta_K g_{\mu \nu}$ in the internal fluctuations of (\ref{eq:gravitonKKExpansion}), the field space metric is seen to be
\begin{equation}\label{eq:Gh}
G^{(h)}_{I_1 J_1}=\frac{1}{4\,V_W}\,\int d^6y \sqrt{\tilde g_6}\,e^{-4A(y)} Y_{I_1}(y)  Y_{J_1}(y)\,,
\end{equation}
which, from the orthogonality relation  (\ref{eq:Ggraviton}), is proportional to the identity matrix. We thus obtain a warped generalization of the usual gravity lagrangian (again taking $\mc M^{kk}$ to be the diagonal part of the metric (\ref{eq:Gh}))
\begin{equation}
\mc L_{kin}^{(h)} = \mc M^{kk} \sum_{I_1}  E^{I_1}_{\mu \nu}(x) \,h_{I_1}^{\mu\nu}(x)
\label{eq:KKGravitonKinetic}
\end{equation}
in terms of the linearized Einstein tensor
\begin{equation}
 E^{I_1}_{\mu \nu}(x):=\frac{1}{2}\big(\Box h^{I_1}_{\mu \nu}-\eta_{\mu\nu} \Box h^{I_1}+\partial_\mu \partial_\nu h^{I_1}-\partial_\mu \partial^\lambda h^{I_1}_{\lambda \nu}-\partial_\nu \partial^\lambda h^{I_1}_{\lambda \mu}+\eta_{\mu\nu} \partial^\lambda \partial^\rho h^{I_1}_{\lambda \rho}\big)\,.
\end{equation}
Since the four dimensional theory has $\mathcal N=1$ supersymmetry, this is the bosonic part of a D-term. In terms of the real vector superfields $H_\mu$ (which contains the graviton and gravitino) and $E_\mu$ (whose $\bar \theta \theta$ component is the Einstein tensor), this D-term is
\begin{equation}\label{eq:Lh2}
 \mc L^{(h)}_D=M^{kk}\,\sum_{I_1}\,\eta^{\mu \nu} E^{I_1}_\mu\,H^{I_1}_\nu\,,
\end{equation}
where we follow the notations of~\cite{weinberg}.

\subsection{K\"ahler potential for internal metric fluctuations}\label{subsec:Kinternal}

At last, we are ready to address the problem of computing the K\"ahler
potential for the internal metric fluctuations. From our previous
considerations, this comes from the second order fluctuation of $\int
R_{(10)}$ generated by the complex structure and K\"ahler moduli
metric perturbations,
\begin{equation}\label{eq:fluctK1}
\delta(ds^2)=2e^{2A} \,\delta A\,\eta^{\mu \nu}dx^\mu dx^\nu+
e^{-2A}\big(-2 \delta A \,\tilde g_{mn}+\delta \tilde g_{mn}\big)dy^m dy^n\,.
\end{equation}
The universal K\"ahler modulus presents additonal subtleties, and its analysis is left for future work.

The full computation is quite long, so we relegate it to the
Appendix. As a summary, the kinetic terms come from
$$
-\frac{1}{2}\int d^4 x\, d^6y \,\sqrt{\tilde g_6}e^{-2A}\,(\delta g)^{mn}\delta G_{mn}\,.
$$
This turns out to be a sum of various pieces containing $\delta \tilde g_{mn}$, $\delta \tilde g$ and $\delta A$. 
After using the constraint Eq.~(\ref{eq:PhysicalGauge1}), surprisingly most of the terms cancel and we are left with
\begin{equation}\label{eq:Seffwarpu}
S_{kin}^{(u)}=\frac{1}{8\kappa_{10}^2}\int d^4x\int d^6y\,\sqrt{\tilde
g_6}\,e^{-4A}\,\delta \tilde g^{mn} \Box\,\delta
\tilde g_{mn}=\frac{1}{2\kappa_{4}^2}\int d^4x\,G^{(u)}_{I_2 \bar J_2}\,u^{I_2}\,\Box \bar u^{J_2}
\end{equation}
where we have used the expansion of Eq. (\ref{eq:internalKKExpansion}) to
write this in terms of a field space metric,
\begin{equation}\label{eq:Gu}
G^{(u)}_{I_2 \bar J_2}=\frac{1}{4V_W}\int d^6y\,\sqrt{\tilde
g_6}\,e^{-4A}\,Y_{I_2,\,mn}(y) \bar Y_{J_2}^{\widetilde{mn}}(y)\,.
\end{equation}
Note that these expressions are not invariant under 6d gauge transformations
$\delta \tilde g_{mn}=\nabla_{(m}\xi_{n)}$, and thus to make them well defined
we need to specify the gauge, which is \eq{PhysicalGauge2}.  We will discuss
this point further in \cite{DT}.

Again,
the metric (\ref{eq:Gu}) is proportional to the orthogonality relation
(\ref{eq:orthon}), so that the kinetic term is diagonal in the different KK levels,
\begin{equation}
\mc L_{kin}^{(u)} = G^{(u)}\sum_{I_2} u^{I_2} \Box \bar{u}^{I_2}\, .
\label{eq:moduliKineticTerm}
\end{equation}

This is one of the main results of our work. If one is only interested
in the propagator of the metric moduli, then even at strong warping it
is consistent to truncate the analysis to the zero KK level. As was
pointed out by~\cite{gid2}, a warp factor does induce mixings between
the moduli and the \emph{unwarped} KK modes, given in
Eq.~(\ref{eq:unweigenvector}). We will argue in section
\ref{sec:masses} that such modes don't represent light four
dimensional excitations. Rather, these are given by the warped
eigenvectors of Eq.~(\ref{eq:InternalEigenmode}), in terms of which
there is no kinetic mixing.

\vskip 1.5mm

Now we can understand better the construction of the complex moduli metric. 
From Eq.~(\ref{eq:Gu}), the field space metric for complex deformations reads
\begin{equation}\label{eq:Gs}
G_{\alpha\bar{\beta}}^{(S)} = \frac{1}{4 V_W} \int d^6y\, \sqrt{\tilde{g}_6}\, e^{-4A} \,\delta_\alpha \tilde{g}_{mn} \,\delta_{\bar \beta} \tilde{g}^{mn}\,,
\end{equation}
and we have changed to a more familiar notation where $\delta_\alpha \tilde g_{mn}$ denotes the wavefunction $Y_{mn}^{I_2}$ associated to a complex structure deformation $S^\alpha$.
This appears at first sight to agree with the form conjectured by~\cite{gid1}, but there are in
fact some crucial differences due to the consistency of the dimensional reduction that one should take into account.
The analysis is based on the gauge choice
\begin{equation}\label{eq:Sconstr1}
 \delta_\alpha g_{\mu m}=0
\end{equation}
and the metric fluctuations must satisfy the constraints
\begin{equation}\label{eq:Sconstr2}
 \delta_\alpha A=\frac{1}{8} \delta_\alpha \tilde g\;,\;\tilde \nabla^n\big(\delta_\alpha \tilde g_{mn}-\frac{1}{2}\tilde
g_{mn}\,\delta_\alpha \tilde g\big)=4 (\partial^{\tilde n}A)\,\delta_\alpha \tilde
g_{mn}\,,
\end{equation}
which follow from the ten dimensional equations $\delta_\alpha
G_{\mu \nu}=\delta_\alpha G_{\mu m}=0$.  Eqs. (\ref{eq:Sconstr1})
and (\ref{eq:Sconstr2}) specify a unique representative
$\delta_\alpha \tilde g_{mn}$ from each class of (diffeomorphism)
equivalent metric fluctuations. Therefore (\ref{eq:Gs}) gives a
well-defined result.

To see how (\ref{eq:Gs}) differs from the standard moduli space metric, 
as in~\cite{candelas} we
write the metric fluctuation in terms of the
$(2,\,1)$ form $\chi$ which is harmonic in the {\it unwarped} metric,
\begin{equation}
\frac{\partial \tilde g_{mn}}{\partial S^\alpha}  = -\frac{1}{||\Omega||^2} \bar{\Omega}_{m}^{\phantom{1}rs}\chi_{\alpha,\, rsn} \, .
\end{equation}
Then
\begin{equation}
\delta_\alpha \tilde{g}_{mn} = \frac{\partial \tilde g_{mn}}{\partial S^\alpha} + \delta_\alpha \tilde{g}_{mn}^{*}
\end{equation}
where 
$\delta_\alpha \tilde{g}_{mn}^{*}$ is determined by
the constraint equations (\ref{eq:Sconstr2}).  

One can check that given a general form for the warp factor, 
\eq{Sconstr2} requires
$\delta_\alpha \tilde{g}_{mn}^*\ne 0$, 
so this extra term leads to further warp corrections
in the metric \eq{Gs},
\begin{equation}
G_{\alpha\bar{\beta}}^{(S)} = \frac{1}{4 V_W} \int e^{-4A}\chi_\alpha \wedge\bar{\chi}_{\bar{\beta}} + 
	\frac{1}{4 V_W} \int d^6y\, \sqrt{\tilde{g}_6}\, e^{-4A}  \frac{\partial \tilde g_{mn}}{\partial S^\alpha}\,\delta_{\bar{\beta}} \tilde{g}^{*mn} + ...
\end{equation}

The upshot is that, unless there are further conspiracies in the 
determination of the warp factor which cause the extra terms to
cancel, the metric on complex structure moduli
space is not \eq{DGMetric} found in \cite{gid1} but instead contains extra
terms. It would be interesting to
compare this result with the moduli space kinetic term
in \cite{10to4}, which was obtained by generalized complex geometry
methods, and to the warping corrections suggested by a different type of analysis for the
universal K\"ahler modulus sector in \cite{burges}.  
We will return to this question in \cite{DT}.

\subsection{Supersymmetry considerations}\label{subsec:Ksyms}

We conclude by interpreting the previous results from the point of
view of the unbroken $\mathcal N=1$ supersymmetry.

In the unwarped case, the massless four dimensional spectrum falls
into the following $\mc N=2$ multiplets~\cite{Grimm}: one gravity
multiplet with matter fields $(g_{\mu \nu}, V^0)$, $h^{(2,1)}$ vector
multiplets $(V^\alpha, S^\alpha)$, $h^{(1,1)}$ hypermultiplets $(v^A,
b^A, c^A, \rho_A)$, and a tensor multiplet $(B_2, C_2, \tau)$. Here,
$V^K$ are space-time gauge fields that come from decomposing $C_4$ in
harmonic 3-forms; $S^\alpha$ and $\rho_A$ are, respectively, the
complex and K\"ahler moduli. The rest of the fields come from
expanding the ten-dimensional 2- and 4-forms as zero forms ($B_2,
C_2$) and 2-forms ($b^A, c^A$) on the internal part.

In the presence of warping, the four dimensional massless spectrum is shown in Table \ref{table:supermult}. The typical warped field space metric is of the form
\begin{equation}
G_{ij}=\frac{1}{V_W} \int d^6y \sqrt{\tilde g_6}\, e^{-4nA(y)}\,\omega_i(y) \omega_ j(y)
\end{equation}
where $\omega_i(y)$ is the wavefunction on the internal space, and the $n$-dependence is given in the table.
\begin{table}\begin{center}\begin{tabular}{|c|c|c|c|}
\hline
$\mc N=1$ multiplet & multiplicity & matter content &
${\rm exp}(-4nA)$ \\
\hline
gravity & 1& $g_{\mu \nu}$& $n=1$ \\
vector & $1+h^{(2,1)}$&$V^K$& $n=0$ \\
chiral & $1+h^{(2,1)}$&$(\tau,\,S^\alpha)$& $n=1$ \\
chiral & $h^{(1,1)}$&$(\rho^A, v^A)$& $n=1$ \\
chiral & $h^{(1,1)}$&$(b^A, c^A)$& $n=0$ \\
tensor & $h^{(1,1)}$&$(B_2, C_2)$& $n=2$ \\
\hline
\end{tabular}\caption{Supermultiplet structure of type IIB supergravity compactified on a warped Calabi-Yau. The warp factor power $n$ refers to the dependence of the field space metric on $e^{-4A}$.
\label{table:supermult}}\end{center}\end{table}

Massive supermultiplets associated to the KK modes also need to be
included. While the supersymmetry description of spin 1 and 2 massive
multiplets is more involved, in our case they can be obtained from
their massless counterparts, via the appropriate super Higgs
mechanism. The outcome is that each KK level exhibits supermultiplets
analogous to the ones in Table \ref{table:supermult}, and the kinetic
terms in the massless and massive case agree. As a consequence, we can
use the same arguments as in the massless case to restrict the
possible derivative terms.

In this way we can understand why, for instance, there
are no kinetic mixings between the internal metric fluctuations (chiral superfields) and the KK gravitons (real vector superfields).
Similarly, the kinetic term for the gauge supermultiplet is an F-term of the spinor superfield $W_\alpha$. This forbids any warp correction in such a term, because $e^{-4A}$ is not holomorphic.

\section{Geometric Masses for KK modes}\label{sec:masses}

In this section we now compute the geometric masses (nonvanishing in
the limit $G_3 \to 0$) for the various KK excitations. Flux-induced
mass terms are discussed in section \ref{sec:fluxmass}. It will be
seen that such terms do not induce additional mixings between the zero
modes and their KK excitations, precisely due to the orthogonality
relations. We do find mixings between graviton KK modes and those from
the internal metric, and point out that they may be interpreted as a
warped generalization of a Higgs-type mechanism.

\subsection{Scalar field case}\label{subsec:mass-scalar}

To illustrate the physics behind the choice of the proper mode expansion, 
we now work in detail the case of a ten-dimensional scalar field; the other modes follow a similar pattern. Consider the action with a possible nontrivial potential,\begin{equation}
S=\frac{1}{2\kappa_{10}^2}\int d^{10}x \sqrt{g}\,\Big( \partial_M \phi 
\,\partial^M \phi+V(\phi) \Big)\,.
\end{equation}
Using the ansatz
$$
\phi(x,y)=\sum_i\,\varphi_i(x) Y_i(y)
$$
the dimensionally reduced action becomes
\begin{equation}
S=-\frac{1}{2\kappa_{10}^2}\int d^4x \Big[ (Y_i, e^{-4A}Y_j)\,\varphi_i \Box
\varphi_j+(Y_i, \tilde \nabla^2 Y_j)\,\varphi_i \varphi_j+V(\varphi)\Big]
\end{equation}
where we have introduced the natural inner product on the Calabi-Yau manifold,
\begin{equation}\label{eq:inner}
(f,g):=\int d^6y \sqrt{\tilde g_6}\,f(y) g(y)\,.
\end{equation}
Both operators $e^{-4A}$ and $\tilde \nabla^2$ are self-adjoint with
respect to this product, so that we have a well-defined action.

A preferred basis for $Y_i(y)$ would be the one in which both the field space
metric and mass matrix are simultaneously diagonalized, if possible.
In our case, such functions are given as the eigenvectors of the following
differential problem:
\begin{equation}\label{eq:eigenvector2}
\tilde \nabla^2\,Y_i(y)=e^{-4A(y)}\,\lambda_i^2\,Y_i(y)\,.
\end{equation}
Then the action acquires the desired diagonal form
\begin{equation}
S=-\frac{1}{2\kappa_{10}^2}\int d^4x \Big[ G\,\varphi_i(\Box+\lambda_i^2)
\varphi_i+V(\varphi)\Big]\,.
\end{equation}

One arrives to the same results by requiring that the 4d scalar has a
well-defined mass, $\eta^{\mu \nu} \partial_\mu \partial_\nu
\varphi_i=-\lambda_i^2 \varphi_i$. These are the mass eigenstates in
the limit $V \to 0$. It turns out that Eq.~(\ref{eq:eigenvector2}) has
a nice interpretation as a Schr\"odinger equation for the wavefunction
$Y_i$ with a potential determined by the warp factor~\cite{frey,
tye}. Light warped KK modes correspond to the bound states of such
potential, while the unwarped modes are associated to states whose
interactions are warp factor insensitive in a box of size $V_W$.  The
low energy dynamics contains massless modes (such as the 4d graviton)
and these bound states. One could insist on describing the system with
the unwarped eigenvectors but this would require a very large number
of fields, as seen from the overlap matrix $(Y_i, \mathcal Y_j)$.

\subsection{Dilaton and $p$-form KK modes}

The dilaton is a particular case of the previous discussion. After expanding around $\tau_0$, the mass matrix reads
\begin{equation}\label{eq:Mtau}
M^{(\tau)2}_{I_1 J_1}=\frac{1}{4{\rm Im}\, \tau_0\,V_W}\,\int d^6y \sqrt{\tilde g_6}\,Y_{I_1}(y) \,\tilde \nabla^2 Y_{J_1}(y)=\mc M^{kk}_\tau\,\lambda_{I_1}^2\,.
\end{equation}
Therefore,
\begin{equation}
\mc L^{(\tau)}=\mc M^{kk}_\tau \sum_{I_1} t^{I_1} (\Box+\lambda_{I_1}^2) \bar t^{I_1}\,.
\end{equation}

Similarly, we can write down the mass term for the vector coming from $C_4$,
\begin{equation}
M^{(V)2}_{I J}\,V^{I}\wedge \star_4 V^{J} 
\end{equation}
where
\begin{equation}\label{eq:MV}
M^{(V)2}_{I J}=\frac{1}{4\,V_W}\,\int_M \,e^{4A}\,d\chi_{I}(y)\wedge \star_6 d\, \chi_{J}(y)\,.
\end{equation}
Notice that, while the field space metric for the vector $V_\mu$ is unwarped, the warp factor enters into the mass matrix. Therefore this sector also exhibits light bound states, much as in the scalar field discussion.

\subsection{Mass terms from dimensional reduction}\label{subsec:dimredM}

The KK masses for the metric fluctuations follow from the effective
action
\begin{equation}
S_{eff}=-\frac{1}{4\kappa_{10}^2} \int d^{10} x\,\sqrt g\,\Big[(\delta g)^{\mu
\nu}
\delta G_{\mu \nu}+(\delta g)^{mn}
\delta G_{mn}\Big]
\end{equation}
if we consider the metric fluctuations (\ref{eq:metricfluct2}), but with
variations being space-time independent. The conformal rescaling used to explain why there are
no spacetime derivative mixings between the graviton and internal metric modes
does not rule out mass mixings of the form $\delta_K g \,\delta \tilde g$.
Therefore we need to consider both types of fluctuations simultaneously.

The full computation is relegated to Appendix \ref{app:full}. After making use of the constraints in (\ref{eq:PhysicalGauge1}) and (\ref{eq:PhysicalGauge2}), the mass terms simplify to
\begin{equation}\label{eq:Skkmass}
S_{mass}=\frac{1}{4\kappa_{10}^2}\int d^{10}x\sqrt{\tilde g_6}\Big[\frac{1}{2}\delta_K g_{\mu\nu}
\tilde \nabla^2(\delta_K g^{\mu\nu}-\eta^{\mu\nu}\delta_K g)-\delta \tilde g^{mn}
\delta \tilde G_{mn} +\frac{1}{2}\delta_K g \delta \tilde R\Big]\,.
\end{equation}
The first two terms give rise to geometric KK masses for the graviton and
internal metric, while the last one mixes these massive sectors. We conclude that there are no mixing
with the metric moduli, which satisfy $\delta \tilde G_{mn}=0$. It is easily seen that
$\delta \tilde G_{mn}=0$ implies $\delta \tilde{R}=0$ for 
a background unwarped metric which is Ricci-flat, i.e., $\tilde{R}=0$.

Eq.~(\ref{eq:Skkmass}) shows massive gravitons coupled to KK modes
from the internal metric. This has a natural interpretation as a
Higgs-type mechanism triggered by the spontaneous breaking of ten
dimensional diffeomorphism invariance ($\langle g_{\mu \nu} \rangle $
and $\langle g_{mn} \rangle$ are nonzero). For instance, in the
original Kaluza-Klein compactification on $\mathbb R^{(3,1)} \times
S^1$, the infinite tower of massive spin 2 fields comes from combining
the 4d gravitons plus Goldstone modes of spin 0 (from $g_{55}$) and
spin 1 (from $g_{\mu 5}$)~\cite{duff}. As in the gauge theory case, it
should be possible to represent the massive states by a gauge
invariant field combining the states of helicity 0, 1 and 2. This was
done for the $S^1$ case in~\cite{cho}. It would be very interesting to
extend that analysis to the warped compactifications discussed here,
and also to provide an explicitly supersymmetric
construction~\cite{msugra}.

\section{Flux-induced Masses and Mixing}\label{sec:fluxmass}

So far we have analyzed, in turns, how the different degrees of
freedom propagate in spacetime, and then what is the mass structure
that they inherit from the underlying warped geometry. The last missing
piece in the analysis is given by the effect of background fluxes.

In truncations to the zero mode sector, the main role of these quantized
fluxes is to lift the complex moduli, via the Gukov-Vafa-Witten
superpotential. It is well understood how this contribution arises in 
unwarped scenarios but, as expected, the presence of warping introduces many
new subtleties. For instance, at constant warping the Chern-Simons term
is topological and does not contribute~\cite{gkp}. However, a warp factor
introduces a nontrivial moduli dependence, from the relation $C_4=e^{4A}d^4 x$.

In~\cite{gid1}, it was argued that the GVW superpotential in the presence of
warping is not modified. However, this analysis did not take into account
the warp factor variation and the CS contribution. On the other hand, the
analysis of~\cite{gid2} was consistent with a GVW-type
superpotential even in the presence of warping. Their approach is based
on a direct ten dimensional analysis, where the potential is identified as the
time component of the Einstein tensor fluctuation. We would like to understand
what is the 4d role of this, and so we present a derivation based on the
compactified effective action (\ref{eq:Seff1}). 
In this section, we will also
analyze the effect of fluxes on KK modes.

\subsection{No mixing with $p$-form modes}\label{subsec:nonmixp}

Before starting our analysis of the flux potential, we show here that
there are no mass mixings between the complex moduli $S$ and KK modes
coming from $(B_2, C_2, C_4)$. After dimensional reduction, these 10d
forms give 4d forms of various ranks. The first point to note is that,
due to Lorentz invariance, the scalar field $S$ can only mix with the
zero forms; hence we restrict our attention to them.

First consider possible mixings coming from $C_4$ and the self-dual term. To account for self-duality, we set $\tilde F_5=dC_4$ and 
multiply by two the terms where $C_4$ appears. After eliminating half of the degrees of freedom, the 
remaining KK modes from $C_4$ which contribute to $\tilde{F}_5^2$ term
are either 1 and 2-forms in space-time, which cannot lead to mixing with moduli by Lorentz invariance, or 0-forms.  Explicit computation
shows that the scalar coming from $C_4$ does not lead to mixing.

Next, the bilinear terms involving $S$ and the zero forms from $(B_2, C_2)$ come from combining the $|G_3|^2$ and CS terms, yielding the usual term
\begin{equation}\label{eq:pmix}
S_{mix}=-\frac{1}{4\kappa_{10}^2\,{\rm Im}\,\tau}\,\int G_3 \wedge \big(\star_{10}\bar G_3+i C_4 \wedge \bar G_3 \big)\,.
\end{equation}
Here we assume a constant dilaton background; mixings with the dilaton KK modes will be analyzed momentarily. Expanding in a complete basis of internal two forms $\omega_A(y)$, the KK mode contribution to the 3-form reads
$$
\delta G_3^{KK}=d \Big( [c_A(x)-\tau b_A(x)]\,\omega_A(y) \Big)\,,
$$
where a sum over $A$ is implicit. If $\omega_A\,\in H^2(M)$, we recover the usual four dimensional zero modes which do not mix with $S$. Here we are interested in the massive modes, for which $\omega_A$ is not closed. Replacing in (\ref{eq:pmix}) and expanding to quadratic order in 
the fields, we have (note that there are no quadratic terms with spacetime derivatives),
\begin{equation}\label{eq:pmix2}
S_{mix}=-\frac{1}{4\kappa_{10}^2\,{\rm Im}\,\tau}\,\int d^4 x\,(c_A-\tau b_A) \bar S\,\int_M d\omega_A(x)\wedge\partial_{\bar S}\Big(e^{4A}[\star_6 \bar G_3+i \bar G_3] \Big)\,.
\end{equation}
Under a complex moduli fluctuation, the $G_3$ equation of motion implies that $\bar \Lambda=e^{4A}[\star_6 \bar G_3+i \bar G_3]$ is closed, 
so $S_{mix}=0$ after integrating by parts.\footnote{In particular, $\bar{\Lambda}$ is a linear combination of $\Omega$ and $\bar\chi_S$.}

\subsection{Flux-induced mass terms}\label{subsec:fluxmass}

The flux-induced masses for metric moduli and KK modes follow from Eq. (\ref{eq:Seff1}). This involves computing the fluctuated energy momentum tensors from $G_3$ and $F_5$, and then contracting with the fluctuated metric. Also, recalling that we are working in backgrounds satisfying $e^{4A}=\alpha$, one gets extra pieces coming from the gravitational part $\delta G_{MN}$, which depends on the warp factor. Furthermore, following section \ref{subsec:comps}, the equation of motion for $\alpha$ has to be imposed as a constraint, and this introduces flux dependence.

It turns out that the computation may be done including the moduli (the relevant ones here are the complex moduli and axio-dilaton) and their KK modes, in a symmetric way; refer to the Appendix for more details. In summary, the flux induced mass terms including moduli and KK modes are
\begin{eqnarray}\label{eq:Seffwarpflux}
S_{flux}&=&-\frac{1}{2\kappa_{10}^2}\int d^4x\int
d^6y\,\sqrt{\tilde g_6}\,\,e^{-2A}\,\Big\{
|\delta_K\tau|^2 \partial_\tau\partial_{\bar{\tau}} (\frac{G_3\cdot \bar{G}_3}{24\Imtau})+\nonumber\\
&+&\delta \tilde g^{\tilde n}_m \delta \left[\frac{1}{8  \Imtau} \left(G_{npq} \,\bar G^{mpq}- \frac{1}{6} \delta^m_n\,|G_3|^2\right)\right] \Big\}
\end{eqnarray}
where the variation `$\delta$' in the last line includes both the axio-dilaton and internal metric fluctuations. To make the result more compact, indices with tildes are raised with $\tilde g_{mn}$, while the ones without tildes are raised with $g_{mn}=e^{-2A}\tilde g_{mn}$.

Restricting to the zero mode sector, this result shows the usual lifting of the moduli by fluxes. However, we would like 
to stress that we are including KK modes as well, as can be seen by inserting the mode expansion (\ref{eq:internalKKExpansion}) 
into (\ref{eq:Seffwarpflux}). The general analysis of the Appendix shows that there are no flux masses for the graviton 
KK modes. Further, one may check that only the traceless parts of the metric fluctuations are lifted by the fluxes. This 
is the familiar statement that K\"ahler moduli are not stabilized at this level, but it also implies that the trace part 
of the massive modes does not couple to the fluxes.

\subsection{Computation of the potential to all orders}\label{subsec:allV}

One very interesting consequence of Eq. (\ref{eq:Seffwarpflux}) is that the flux contribution may mix the zero modes with the massive fluctuations. It is very important to understand such mixings, since so far all the other terms in our effective action do not exhibit this effect (at least to quadratic order).

Unfortunately, $S_{flux}$ presents a rather complicated structure and it seems that statements about mixings will depend strongly on the particular background, with the corresponding flux choice and form of $\delta \tilde g_{mn}$. Nevertheless, we now describe an alternative approach for finding $S_{flux}$ which may be better suited for answering these sorts of questions.

The method is based on two observations: first, to compute the potential it is enough to consider space-time independent fluctuations. Also, the expression as a power series (\ref{eq:Seff1}) is only necessary to identify the `geometrical' KK masses. In order to find the flux potential such terms may be set to zero, and an appropriate use of the 10d equations of motion gives us an answer to all orders in the fluctuations. 

This is in fact the spirit of the original GKP derivation~\cite{gkp} or the more detailed approach of~\cite{gid1}. However, for a nontrivial warp factor some terms would be missing in their derivation, and we also want to include KK modes. 

The terms contributing to the potential are
\begin{equation}\label{eq:Sfluxall1}
S_{flux}=\frac{1}{2\kappa_{10}^2} \int
d^{10}x\,\sqrt{-g}\, \Big\{R -\frac{G_3 \cdot \bar G_3}{12 {\rm Im
\tau}} -\frac{\tilde F_5^2}{480}\Big\}
-\frac{i}{8\kappa_{10}^2} \int \frac{C_4 \wedge G_3 \wedge \bar
G_3}{{\rm Im \tau}}\,.
\end{equation}
First, the Ricci scalar part has the form
\begin{equation}\label{eq:Sr}
\int d^{10}x \sqrt{-g} R = \int d^4x  \int d^6y
\,\sqrt{g_6}\,\left[-8 e^{4A}(\nabla A)^2 +\ldots\right]\,,
\end{equation}
where the dots refer to terms induced by the KK modes, which are related to their geometric masses and do not depend on moduli. The flux dependence here comes from the equation of motion 

Next, the $G_3$ term is already in the desired form. Finally, after integrating by parts and using the Bianchi identity for the 5-form, the $\tilde F_5^2$ and CS terms give
$$
 \int
d^{10}x\,\sqrt{-g}\, \frac{\tilde F_5^2}{480}
+\frac{i}{4} \int \frac{C_4 \wedge G_3 \wedge \bar
G_3}{{\rm Im \tau}}\to \frac{i}{4}\int d^4x\,\int \frac{e^{4A(y)}}{{\rm
Im}\,\tau}\,G_3 \wedge \bar G_3\,.
$$
Combining these contributions, we arrive to
\begin{equation}\label{eq:Sv2}
S_{flux}=-\frac{1}{4\kappa_{10}^2}\int d^4x \int \frac{e^{4A}}{{\rm Im}\,\tau}\, G_3
\wedge \big( \star_6 \bar G_3+i \bar G_3 \big)\,.
\end{equation}
As a check, the second order variation of this expression reproduces our previous result Eq. (\ref{eq:Seffwarpflux}).

Summarizing, Eq. (\ref{eq:Sv2}) gives the full flux potential for the
metric fluctuations including KK modes. This has the same form as the
potential including only complex moduli. There are, however, two
differences. This expression is valid including axio-dilaton
fluctuations, while the original derivation set $\tau$ to a
constant. Further, the massive metric modes are encoded in the Hodge
star. One cannot use the method of~\cite{gid1} to obtain the GVW
superpotential from here, since for arbitrary massive fluctuations we
do not know which are the 3-forms with definite self-duality
properties under $\star_6$. We plan to analyze these issues in a
future work.

\vskip 2mm


\subsection*{Acknowledgements}

We thank F.~Denef, R.~Essig, A.~Frey, S.~Giddings, J.~Gray, T.~Grimm,
K.~Intriligator, S.~Kachru, S.~Klevtsov, H.~Lu, A.~Nacif, P.~Ouyang,
M.~Porrati, K.~van den Broek, T.~van Riet and K.~Sinha for helpful discussions
and comments. G.S. and B.U. are supported in part by NSF CAREER Award
No. PHY-0348093, DOE grant DE-FG-02-95ER40896, a Research Innovation
Award and a Cottrell Scholar Award from Research Corporation. 
M.R.D. is supported in part by DOE grant DE-FG02-96ER40959.
G.T. is supported as a Research Assistant in the Rutgers Department of
Physics.

\appendix


\section{Axial gauge equations for time dependent moduli}\label{app:flux}

As discussed in Section 2.1, when
the internal metric moduli are space-time dependent there can be additional
fluctuations of the 10-dimensional metric which are proportional to spacetime
derivatives of the moduli~\cite{gid2}
\begin{equation}\label{eq:compmetric}
\delta_c ds^2=2\partial_\mu \partial_\nu u^I(x) e^{2A} K_I(y) dx^\mu dx^\nu + 2e^{2A}B_{Im}(y) \partial_\mu u^I(x) dx^\mu dy^m\;.
\end{equation}
In general, these fluctuations can be gauged away by an appropriate gauge choice
\begin{eqnarray}
\epsilon_\mu &=&  - \partial_\mu u^I(x) e^{2A} K_I(y)\\
\epsilon_m &=& u^I(x) \partial_m \big(e^{2A} K_I(y)\big) - 2 e^{2A} B_{Im}(y) u^I(x)\, .
\end{eqnarray}
The only remaining gauge transformations allowed that preserve this gauge choice are spacetime independent
internal diffeomorphisms $\partial_\mu \epsilon_m = 0$ and pure four-dimensional diffeomorphisms $\partial_m \epsilon_\mu = 0$.
Note that this gauge transformation induces a non-zero time dependent transformation of the fluctuated internal metric,
$\delta\tilde{g}_{mn} \rightarrow \delta\tilde{g}_{mn} + 2 \nabla_{(m} \epsilon_{n)}$, so that if the spacetime dependent fluctuation started
in transverse traceless gauge, it will no longer remain so.  

One of the special features about GKP backgrounds that simplified our calculations is that the flux 
sector does not get modified when the moduli are promoted to space-time dependent fields. 
Therefore it's worth describing this in detail.

Promoting the moduli to spacetime dependent fields leads to the possibility of fluctuations in the $p$-form fluxes of the form~\cite{gid2}
\begin{equation}\label{eq:pformcomp}
\delta_c(C_2-\tau B_2)=du^I \wedge T_I\;,\;\delta_c C_4=du^I \wedge S_I^{(3)}+\star_4 du^I \wedge S_I\,.
\end{equation}
These fluctuations are found as solutions to the fluctuated equations of motion
\begin{equation}\label{eq:constr1}
d\big[\delta_I(\star_{10} G_3 -i C_4 \wedge G_3) \big]=0\,,
\end{equation}
\begin{equation}\label{eq:constr2}
d\delta_I \tilde{F}_5 = \delta_I \left(H_3\wedge F_3\right) + 2
\kappa_{10}^2 T_{D3} \delta_I (\rho_3^{loc})\, ,
\end{equation}
where $\delta_I:=u^I(x) \partial_I$. The fluctuations are subject to the self-duality relation $\delta\tilde{F}_5 = \delta (\star_{10}\tilde{F}_5)$. 

We will take the localized sources to be far away from the cycles on
which the moduli are localized, so that
$\delta(\rho_3^{loc}) \approx 0$. For instance, for a D3 brane at a distance $y=y_0$ from a warped throat with scale $e^{A_{min}} \sim \Lambda$, these corrections are suppressed by $\Lambda^4/|y_0|^4$. More concretely, in the embedding of the KS solution in a compact 3-fold given in~\cite{gkp}, $|y_0|^6 \sim V_{(6)}$. However, it would also be interesting to understand the effects of localized sources inside the throat, for applications to sequestering and supersymmetry breaking (see, for instance,~\cite{kpv, Sequestering, aharony}).

Solving the flux equations of motion (\ref{eq:constr1}, \ref{eq:constr2}), subject to the self-duality
constraint, it is easy to see that the internal space part of the fluctuations (\ref{eq:pformcomp})
are constant, $d T_I = dS_I = 0$ (we also find that $S^{(3)}_I = 0$).  Since these are closed one forms
on a Calabi-Yau space, they must be exact, which implies that they can be gauged away. This is the gauge choice we have used in this work.

There are situations where this is not the case, and having time-dependent moduli has physical effects on the $p$-forms; a related example is analyzed in~\cite{gray}. Here, the reason why we can set $T_I = S_I = S_I^{(3)} = 0$ is that the equations of motion (\ref{eq:constr1}) and (\ref{eq:constr2}) are actually \emph{time-independent}. Therefore, promoting the moduli to fields does not induce changes in this sector.

\section{Warped KK mode effective action}
\label{app:full}

\subsection*{Fluctuated equations of motion}\label{subsec:flucteom}

We begin by collecting the relevant supergravity formulas
for the ten dimensional fluctuations, following~\cite{gid2}.
The fluctuated Einstein's equations are
\begin{equation}
\delta G^M_N=\kappa_{10}^2 \delta T^M_N\,,
\end{equation}
where raising one of the indices simplifies the computation of
(\ref{eq:fluctR}) by eliminating derivatives of the warp factor.
First,
\begin{eqnarray}\label{eq:Gmunu}
\delta (G^\mu_\nu)&=& \delta^\mu_\nu \delta \big[-\frac{1}{2}e^{2A}
\tilde R+\frac{3}{4} e^{-6A} \partial_m e^{4A} \partial^{\tilde{m}} e^{4A} - \frac{1}{2} e^{-2A} \tilde{\nabla}^2 e^{4A}
\big]+\nonumber\\
&+&e^{-2A} \delta_K G^{(4)
\hat{\mu}}_{\phantom{11}\nu}-\frac{1}{2}e^{2A} \tilde
\nabla^2(\delta_K g^{\hat\mu}_\nu-\delta^\mu_\nu \delta_K g)
\end{eqnarray}
where $\delta_K g:=\eta^{\mu \nu}\delta_K g_{\mu \nu}$ and $\delta_K
G^{(4)}$ is the four-dimensional Einstein's tensor for $\eta_{\mu
\nu}+\delta_K g_{\mu \nu}$. On the other hand,
\begin{eqnarray}\label{eq:Gmn}
\delta (G^m_n)&=& \delta \big[e^{2A}(\tilde G^{\tilde
m}_n+
\frac{1}{4} e^{-8A} \delta^{\tilde{m}}_n \partial_p e^{4A}\partial^{\tilde{p}} e^{4A} 
-\frac{1}{2} e^{-8A} \partial_n e^{4A} \partial^{\tilde{m}} e^{4A})\big] \nonumber\\
&+&\frac{1}{2}\big[-e^{-2A} \tilde \nabla^{\tilde (m}(e^{4A}
\partial_{n)} \delta_K g)+\delta^m_n \tilde \nabla^{\tilde p}(e^{2A}
\partial_p \delta_K g)
\big]+\nonumber\\
&-&\frac{1}{2} \delta^m_n e^{-2A} \delta_K R^{(4)} - \frac{1}{2} e^{-2A} \Box \delta\tilde{g}^{\tilde{m}}_n + 
\frac{1}{4} \delta^m_n e^{-2A} \Box \delta\tilde{g}\,.
\end{eqnarray}
The index notation is $\delta_K g^{\hat \mu}_\nu:=\eta^{\mu \lambda}\,\delta_K
g_{\lambda \nu}$ and $\delta \tilde g^{\tilde m}_n:=\tilde
g^{mp}\,\delta \tilde g_{pn}$, and similarly for other tensors.
Finally,
\begin{equation}\label{eq:Gmun}
\delta G^\mu_m=e^{-2A} \partial^{\hat
\mu}\big(-\frac{1}{4}\partial_m \delta \tilde g-2 \partial^{\tilde
p}A\,\delta \tilde g_{mp}+\frac{1}{2}\,\tilde \nabla^{\tilde
p}\,\delta \tilde g_{mp} \big)\,.
\end{equation}

The energy momentum tensor has contributions from the three and five
forms, and from other local sources (3-branes and orientifolds):
$T_{MN}=T_{MN}^{(5)}+T_{MN}^{(3)}+T_{MN}^{loc}$. A straightforward
computation gives
\begin{eqnarray}\label{eq:TMN}
T_\nu^{(5)\mu}+T_\nu^{(3)\mu}&=&-\frac{1}{4\kappa_{10}^2}\,\delta_\nu^\mu\big(
e^{-6A}\,\partial_m \alpha \partial^{\tilde m}\alpha+\frac{G_3 \cdot
\bar G_3}{8{\rm Im}\,\tau}\big)\,,\\
T_m^{(5)n}+T_m^{(3)n}&=&\frac{1}{2\kappa_{10}^2}\,\big[-e^{-6A}
(\partial_m \alpha \partial^{\tilde n}
\alpha-\frac{1}{2}\delta^n_m\partial_p \alpha
\partial^{\tilde p} \alpha)+\nonumber\\
&+&\frac{1}{4{\rm Im}\,\tau}(G_{mpq} \bar
G^{npq}+G^n_{\phantom{1}pq} \bar G_m^{\phantom{1}pq}-\frac{1}{3}
\delta^n_m G_3 \cdot \bar G_3) \big]\,.
\end{eqnarray}

Besides the fluctuated Einstein's equations, the three and five form
equations of motion become \emph{constraints} on the four
dimensional fluctuations:
\begin{equation}\label{eq:fluctF5}
\delta\big(\tilde \nabla^2 \alpha-2e^{-6A}\partial_m \alpha
\partial^{\tilde m} e^{4A} \big)=\delta \big(i e^{2A}\frac{G_{mnp}\,(\star_6 \bar G)^{mnp}}{12\,
{\rm Im}\,\tau}+2\kappa_{10}^2e^{2A}\,T_3 \rho_3^{loc} \big)
\end{equation}
\begin{equation}\label{eq:fluctG3}
d[e^{4A}\,(\delta\,\star_6 G_3)]=0
\end{equation}
where in the last line the fact that the background is ISD was used.
Another constraint follows from fluctuating the self-duality
condition $\star_{10} \tilde F_5=\tilde F_5$, which is equivalent to
$\star_{10}\,dC_4=B_2 \wedge F_3$. In our case, under a metric
fluctuation, fluxes stay fixed at their quantized values, so we get
\begin{equation}\label{eq:selfdualfluct}
\delta[\star_{10}\,dC_4]=0\,.
\end{equation}
In this work we will restrict to fluctuations that preserve the form of the BPS condition $\alpha=e^{4A}$, i.e., $\delta(\alpha-e^{4A})=0$.

These properties of the BPS flux compactifications that we are
considering are at the root of many restrictions in the possible
mixings. The $(\mu\nu)$ fluctuated Einstein
equation simplifies to
\begin{eqnarray}\label{eq:deltamunu}
\delta G^\mu_\nu-\kappa_{10}^2 \delta T^\mu_\nu&=&e^{-2A}\,\delta_K
G^{(4)\,\hat \mu}_\nu-\frac{1}{2}e^{2A}\,\tilde
\nabla^2\big(\delta_K g^{\hat \mu}_\nu-\delta^\mu_\nu\,\delta_K g
\big)+\nonumber\\
&-&\frac{1}{2}\delta^\mu_\nu e^{2A}\,\delta \tilde
R-\delta^\mu_\nu\kappa_{10}^2T_3 \delta
\rho_3^{loc}-\kappa_{10}^2\,\delta T_{loc\,\nu}^\mu\,.
\end{eqnarray}
We remind the reader that hats denote indices raised with $\eta_{\mu
\nu}$. Similarly, in the $(mn)$ component all the $\alpha$ and $A$
variations cancel, yielding
\begin{eqnarray}\label{eq:deltamn}
\delta G^m_n-\kappa_{10}^2 \delta T^m_n&=&e^{2A}\,\delta\,\tilde
G^{\tilde m}_n-\delta\big[\frac{1}{8{\rm Im}\,\tau}\big(2G_{mpq} \bar
G^{npq}-\frac{1}{3}
\delta^n_m G_3 \cdot \bar G_3 \big)\big]+\nonumber\\
 &-&\frac{1}{2}e^{-2A}
\Box \,\delta \tilde g^{\tilde m}_n+\frac{1}{4}e^{-2A}
\delta^m_n\,\Box\, \delta \tilde g-\frac{1}{2} e^{-2A}\,\tilde
\nabla^{\tilde m} (e^{4A}
\partial_n \delta_K g)+\nonumber\\
&+&\frac{1}{2}\delta^m_n \tilde \nabla^{\tilde p}(e^{2A}\partial_p
\delta_K g)-\frac{1}{2}e^{-2A}\delta^m_n\delta_K R^{(4)\,\hat
\mu}_\mu-\kappa_{10}^2 \delta T_{loc\,n}^m
\end{eqnarray}
where we already combined the symmetric terms, since this is going
to be contracted with $\delta g_{mn}$; tildes refer to indices
raised with $\tilde g_{mn}$. Finally, the $(\mu m)$ equation gives the constraint
\begin{equation}\label{eq:Gmuncons}
-\frac{1}{4}\partial_m\delta \tilde g-2 \partial^{\tilde p}A\,\delta
\tilde g_{mp}+\frac{1}{2}\,\tilde \nabla^{\tilde p}\,\delta \tilde
g_{mp} =0\,.
\end{equation}

\subsection*{Derivation of the Effective Action}\label{subsec:dynamicskk1}

Using the previous fluctuated equations of motion gives, after contracting (\ref{eq:metricfluct2}) with (\ref{eq:Gmunu}) and (\ref{eq:Gmn}),
\begin{eqnarray}\label{eq:Seffgeneral}
S_{eff}&=&-\frac{1}{2}\int d^4x\int d^6y\,\sqrt{\tilde g_6}\Big[
e^{-4A}\,\delta_K g_\mu^{\hat \nu} \,\delta_K G^{(4)\,
\hat{\mu}}_\nu-\frac{1}{2}\delta_K g_\mu^{\hat \nu}\tilde
\nabla^2\big(\delta_K g^{\hat
\mu}_\nu-\delta^\mu_\nu\,\delta_K g \big)+\nonumber\\
&-&\frac{1}{2}e^{-4A}\,\delta \tilde g^{\tilde n}_m \Box\,\delta
\tilde g^{\tilde m}_n+\delta \tilde g^{\tilde n}_m\,\delta \tilde
G^{\tilde m}_n-\frac{e^{-2A}}{8\,{\rm Im}\,\tau}\delta \tilde
g^{\tilde n}_m \big(2 G_{npq} \,\delta \bar G^{mpq}
-\frac{1}{3}\delta^m_n\,\delta [G_3 \cdot \bar
G_3]\big)\nonumber\\
&+& \frac{1}{2}(-\delta_K g\, \delta \tilde R+\frac{3}{4} \delta_K g
\,\tilde \nabla^2 \delta \tilde
g)+\frac{1}{2}e^{-2A}\big\{-e^{-2A}\delta \tilde g^{\tilde n}_m
\tilde \nabla^{\tilde m}(e^{4A}\partial_n \delta_K g)+\nonumber\\
&-&\frac{1}{2}\delta \tilde g \tilde \nabla^{\tilde p}(e^{2A}
\partial_p \delta_K g)+\frac{1}{4}e^{-2A} \delta \tilde g \tilde \nabla^{\tilde
m}(e^{4A}\partial_m \delta_K g) \big\}-\kappa_{10}^2\,e^{-2A}\delta
g\, \delta T_{loc} \Big]
\end{eqnarray}
where the last term stands for
\begin{equation}\label{eq:deltaloc}
\delta g\, \delta T_{loc}:=T_3\,\delta_K g \, \delta
\rho_3^{loc}+\delta_K g_{\mu}^{\hat \nu}\,\delta
T_{loc\,\nu}^\mu-\frac{1}{4} \delta \tilde g \,\delta
T_{loc\,m}^m+\delta \tilde g^{\tilde n}_m \,\delta T_{loc\,n}^m\,.
\end{equation}
To avoid cluttering, the dilaton fluctuations have not been included; they will be considered shortly.

The final step in the computation is to impose the constraint  (\ref{eq:Gmuncons}), after which we obtain
\begin{equation}\label{eq:Seffwarp1}
S_{eff}=S_{eff}^{(KK)}+S_{eff}^{(S)}+S_{eff}^{(\tau)}+S_{mix}
\end{equation}
where
\begin{equation}\label{eq:SeffwarpKK}
S_{eff}^{(KK)}=-\frac{1}{2}\int d^4x\int d^6y\,\sqrt{\tilde
g_6}\big[e^{-4A}\,\delta_K g_\mu^\nu \,\delta_K G^{(4)\,
\mu}_\nu-\frac{1}{2}\delta_K g_\mu^\nu \tilde \nabla^2\big(\delta_K
g^\mu_\nu-\delta^\mu_\nu\,\delta_K g \big)\big]
\end{equation}
is the usual graviton KK mode action, while the one for internal
fluctuations is
\begin{equation}\label{eq:SeffwarpS}
S_{eff}^{(S)}=-\frac{1}{2}\int d^4x\int d^6y\,\sqrt{\tilde
g_6}\,\big[-\frac{1}{2}e^{-4A}\,\delta \tilde g^{mn} \Box\,\delta
\tilde g_{mn}+\delta \tilde g^{mn}\,\delta \tilde G_{mn} \big]
\end{equation}
and the dilaton part reads
\begin{equation}\label{eq:Seffwarptau}
S_{eff}^{(\tau)} = \int d^4x \int d^6y \sqrt{\tilde{g}_6} \big[\frac{1}{4(\Imtau_0)^2}e^{-4A} \delta_K\bar{\tau} \Box \delta_K\tau + \frac{1}{4(\Imtau_0)^2}\delta_K\bar{\tau} \tilde{\nabla}^2 \delta_K\tau \big]\,.
\end{equation}
The mixing term has the structure
\begin{equation}\label{eq:Seffwarpmix}
S_{mix}=\frac{1}{2}\int d^4x\int d^6y\,\sqrt{\tilde g_6}\,
\Big[\frac{1}{2}\delta_K g \,\delta \tilde
R+\kappa_{10}^2\,e^{-2A}\delta g\, \delta T_{loc} \Big]+S_{flux}\,,
\end{equation}
where the flux contribution is given by
\begin{eqnarray}\label{eq:appSeffwarpflux}
S_{flux}&=&-\int d^4x\int
d^6y\,\sqrt{\tilde g_6}\,\,e^{-2A}\,\Big\{
|\delta_K\tau|^2 \partial_\tau\partial_{\bar{\tau}} (\frac{G_3\cdot \bar{G}_3}{24\Imtau})+\nonumber\\
&+&\delta \tilde g^{\tilde n}_m \delta \left[\frac{1}{8  \Imtau} \left(G_{npq} \,\bar G^{mpq}- \frac{1}{6} \delta^m_n\,|G_3|^2\right)\right] \Big\}
\end{eqnarray}
Where there is no confusion we have eliminated the hats and tildes, and in the last line the variation $\delta$ includes complex moduli and dilaton fluctuations. The flux contribution may in principle mix the complex structure
zero modes with their KK excitations and the axio-dilaton KK modes.

\section{Summary of warped effective theory}
\label{app:4dReview}

Inserting the mode expansions of the fields into (\ref{eq:Seffwarp1}) and integrating over the internal
space gives rise to the complete four dimensional effective action.  Throughout the paper we computed
the effective action in separate sections.
Here we collect our expressions for the four dimensional effective Lagrangian for the complex and K\"ahler
moduli, graviton KK modes, and dilaton KK modes:
\begin{eqnarray}
{\mathcal L}_{eff} &=& {\mathcal L}_{u} + {\mathcal L}_{KK} + {\mathcal L}_\tau + {\mathcal L}_{mix} \nonumber \\
	&=&  G^{(u)}\sum_{I_2} \,\bar{u}^{I_2}\left(\Box + \lambda_{I_2}^2 \right)u^{I_2} + {\mathcal M}^{kk} \sum_{I_1}\,h^{\mu\nu}_{I_1}\left( E_{\mu\nu}^{I_1} + \lambda_{I_1}^2 h_{\mu\nu}^{I_1} \right) \nonumber \\
&& + {\mathcal M}^{kk}_\tau \sum_{I_1}\,\bar{t}^{I_1} \left( \Box +\lambda_{I_1}^2 \right)t^{I_1} 
	+ \sum_{I_1,J_1} A_{I_1J_1} \bar{t}^{I_1} t^{J_1} - \sum_{I_1 J_2} \gamma_{I_1J_2} \bar{u}^{J_2} h^{I_1}\nonumber\\ &&+ \sum_{I_2,J_2} (\alpha_{I_2 J_2}-\beta_{I_2 J_2})\,\bar{u}^{I_2} u^{J_2}  + \sum_{I_2,J_1} B_{I_2J_1}\,u^{I_2} \bar{t}^{J_1} \,.
\end{eqnarray}
The various matrices here are defined as
\begin{eqnarray}
G^{(u)} &=& \frac{1}{4V_W} \int d^6y \sqrt{\tilde{g}_6} e^{-4A} \,Y^{I_2}_{mn} Y^{I_2\widetilde{mn}}  \\
{\mathcal M}^{kk} &=&-4 (\Imtau_0)^2\,{\mathcal M}^{kk}_\tau= \frac{1}{2V_W} \int d^6y \sqrt{\tilde{g}_6} e^{-4A}\, Y^{I_1} Y^{I_1}\\
\alpha_{I_2 J_2} &=& \frac{3}{8\mbox{Im}\tau V_W} \int d^6y \sqrt{\tilde{g}_6} \,Y_{mn}^{I_2} (Y^{J_2\widetilde{mm'}}-\frac{1}{4} Y^{J_2\tilde{p}}_p \tilde{g}^{mm'}) G^n_{pq} \bar{G}_{m'}^{pq}
\end{eqnarray}
\begin{eqnarray}
\beta_{I_2J_2} &=& \frac{1}{16\mbox{Im}\tau V_W} \int d^6y \sqrt{\tilde{g}_6} \,Y^{I_2} (Y^{J_2\widetilde{rr'}}-\frac{1}{4} Y^{J_2\tilde{p}}_p \tilde{g}^{rr'}) G_{pqr}\bar{G}^{pq}_{r'} \\
\gamma_{I_1 J_2} &=& \frac{1}{4V_W} \int d^6y \sqrt{\tilde{g}_6}\, Y^{I_1}  \tilde \nabla^m \tilde \nabla^n Y^{I_2}_{mn} \\
A_{I_1J_1} &=& -\frac{1}{48V_W}\int d^6y \sqrt{\tilde{g}_6}~ e^{-2A}  \frac{|G_3|^2}{(\Imtau_0)^3}\,Y^{I_1} Y^{J_1}\\
B_{I_2J_1} &=&\frac{1}{4 V_W} \int d^6y\sqrt{\tilde{g}_6} e^{-2A} Y^{I_2 \widetilde{mn}} \bar{Y}^{J_1}  \nonumber \\
		&& \left[ \frac{G_{mpq}\bar{G}_n^{pq} - \frac{1}{3} g_{mn}|G_3|^2}{(\Imtau_0)^2} + \frac{1}{\Imtau_0} (\tau H_{mpq}H_n^{pq}-F_{mpq}H_n^{pq})\right]
\end{eqnarray}
In the limit of weak warping $e^A\approx 1$, this reduces to the usual unwarped
4-dimensional effective theory.


\begin{thebibliography}{99}

\bibitem{ads}J.~M.~Maldacena,
  Adv.\ Theor.\ Math.\ Phys.\  {\bf 2}, 231 (1998)
  [Int.\ J.\ Theor.\ Phys.\  {\bf 38}, 1113 (1999)]
  [arXiv:hep-th/9711200].

\bibitem{dv} F.~Cachazo, K.~A.~Intriligator and C.~Vafa,
  Nucl.\ Phys.\  B {\bf 603}, 3 (2001)
  [arXiv:hep-th/0103067].
R.~Dijkgraaf and C.~Vafa,
  arXiv:hep-th/0208048.

\bibitem{ks} I.~R.~Klebanov and M.~J.~Strassler,
  JHEP {\bf 0008}, 052 (2000)
  [arXiv:hep-th/0007191].

\bibitem{italian}
  A.~Ceresole, G.~Dall'Agata, R.~D'Auria and S.~Ferrara,
  Phys.\ Rev.\  D {\bf 61}, 066001 (2000)
  [arXiv:hep-th/9905226].
 A.~Ceresole, G.~Dall'Agata and R.~D'Auria,
  JHEP {\bf 9911}, 009 (1999)
  [arXiv:hep-th/9907216].

\bibitem{ksfluct1} M.~Berg, M.~Haack and W.~Mueck,
  Nucl.\ Phys.\  B {\bf 736}, 82 (2006)
  [arXiv:hep-th/0507285].
  M.~Berg, M.~Haack and W.~Mueck,
  Nucl.\ Phys.\  B {\bf 789}, 1 (2008)
  [arXiv:hep-th/0612224].

\bibitem{ksfluct2}
A.~Dymarsky and D.~Melnikov, 
  arXiv:0710.4517 [hep-th].
M.~K.~Benna, A.~Dymarsky, I.~R.~Klebanov and A.~Solovyov,
  arXiv:0712.4404 [hep-th].
  M.~K.~Benna, A.~Dymarsky and I.~R.~Klebanov,
  JHEP {\bf 0708}, 034 (2007)
  [arXiv:hep-th/0612136].

\bibitem{Gubser:2004qj}
  S.~S.~Gubser, C.~P.~Herzog and I.~R.~Klebanov,
  JHEP {\bf 0409}, 036 (2004)
  [arXiv:hep-th/0405282].

\bibitem{kpv} S.~Kachru, J.~Pearson and H.~L.~Verlinde,
  JHEP {\bf 0206}, 021 (2002)
  [arXiv:hep-th/0112197].
  O.~DeWolfe, S.~Kachru and M.~Mulligan,
  arXiv:0801.1520 [hep-th].

\bibitem{ag}M.~Aganagic, C.~Beem, J.~Seo and C.~Vafa,
  Nucl.\ Phys.\  B {\bf 789}, 382 (2008)
  [arXiv:hep-th/0610249].
  J.~J.~Heckman, J.~Seo and C.~Vafa,
  JHEP {\bf 0707}, 073 (2007)
  [arXiv:hep-th/0702077].

\bibitem{dst} M.~R.~Douglas, J.~Shelton and G.~Torroba,
  arXiv:0704.4001 [hep-th].
  
\bibitem{obsgeom}  
S.~Franco and A.~M.~Uranga,
  JHEP {\bf 0606}, 031 (2006)
  [arXiv:hep-th/0604136].
S.~Franco, A.~Hanany, F.~Saad and A.~M.~Uranga,
  JHEP {\bf 0601}, 011 (2006)
  [arXiv:hep-th/0505040].
S.~Franco, I.~Garcia-Etxebarria and A.~M.~Uranga,
  JHEP {\bf 0701}, 085 (2007)
  [arXiv:hep-th/0607218].

\bibitem{obsgeom2}
  D.~E.~Diaconescu, A.~Garcia-Raboso and K.~Sinha,
  JHEP {\bf 0606}, 058 (2006)
  [arXiv:hep-th/0602138].
  D.~Malyshev,
  arXiv:0705.3281 [hep-th].
K.~Sinha,
  arXiv:0709.2932 [hep-th].

\bibitem{iss} K.~Intriligator, N.~Seiberg and D.~Shih,
  JHEP {\bf 0604}, 021 (2006)
  [arXiv:hep-th/0602239].
  
\bibitem{meta1} R.~Argurio, M.~Bertolini, S.~Franco and S.~Kachru,
  JHEP {\bf 0701}, 083 (2007)
  [arXiv:hep-th/0610212].
  R.~Argurio, M.~Bertolini, S.~Franco and S.~Kachru,
  JHEP {\bf 0706}, 017 (2007)
  [arXiv:hep-th/0703236].
\bibitem{meta2} 
 R.~Essig, K.~Sinha and G.~Torroba,
  JHEP {\bf 0709}, 032 (2007)
  [arXiv:0707.0007 [hep-th]].
  M.~Buican, D.~Malyshev and H.~Verlinde,
  arXiv:0710.5519 [hep-th].

\bibitem{intersect} A.~Giveon and D.~Kutasov,
  Nucl.\ Phys.\  B {\bf 778}, 129 (2007)
  [arXiv:hep-th/0703135].
  A.~Giveon and D.~Kutasov,
  JHEP {\bf 0802}, 038 (2008)
  [arXiv:0710.1833 [hep-th]].

\bibitem{broek} M.~Dine, G.~Festuccia, A.~Morisse and K.~van den Broek,
  arXiv:0712.1397 [hep-th].

\bibitem{gkp} S.~B.~Giddings, S.~Kachru and J.~Polchinski,
  Phys.\ Rev.\  D {\bf 66}, 106006 (2002)
  [arXiv:hep-th/0105097].

\bibitem{Grana:2005jc}
  M.~Grana,
  Phys.\ Rept.\  {\bf 423}, 91 (2006)
  [arXiv:hep-th/0509003].

\bibitem{review1} M.~R.~Douglas and S.~Kachru,
  Rev.\ Mod.\ Phys.\  {\bf 79}, 733 (2007)
  [arXiv:hep-th/0610102].
F.~Denef, M.~R.~Douglas and S.~Kachru,
  Ann.\ Rev.\ Nucl.\ Part.\ Sci.\  {\bf 57}, 119 (2007)
  [arXiv:hep-th/0701050].

\bibitem{review2}F.~Denef,
  arXiv:0803.1194 [hep-th].

\bibitem{freyThesis}
A.~R.~Frey,
  arXiv:hep-th/0308156.

\bibitem{becker} K.~Becker and M.~Becker,
  Nucl.\ Phys.\  B {\bf 477}, 155 (1996)
  [arXiv:hep-th/9605053].

\bibitem{verlinde} H.~L.~Verlinde,
  Nucl.\ Phys.\  B {\bf 580}, 264 (2000)
  [arXiv:hep-th/9906182].

\bibitem{sethi} K.~Dasgupta, G.~Rajesh and S.~Sethi,
  JHEP {\bf 9908}, 023 (1999)
  [arXiv:hep-th/9908088].

\bibitem{gid2} S.~B.~Giddings and A.~Maharana,
  Phys.\ Rev.\  D {\bf 73}, 126003 (2006)
  [arXiv:hep-th/0507158].

\bibitem{grimm-hitchin}
  I.~Benmachiche and T.~W.~Grimm,
  Nucl.\ Phys.\  B {\bf 748}, 200 (2006)
  [arXiv:hep-th/0602241].

\bibitem{10to4}
P.~Koerber and L.~Martucci,
  JHEP {\bf 0708}, 059 (2007)
  [arXiv:0707.1038 [hep-th]].


\bibitem{gid1} O.~DeWolfe and S.~B.~Giddings,
  Phys.\ Rev.\  D {\bf 67}, 066008 (2003)
  [arXiv:hep-th/0208123].

\bibitem{FreyPolchinski}
A.~R.~Frey and J.~Polchinski,
  Phys.\ Rev.\  D {\bf 65}, 126009 (2002)
  [arXiv:hep-th/0201029].

\bibitem{alwis}
S.~P.~de Alwis,
  Phys.\ Rev.\  D {\bf 68}, 126001 (2003)
  [arXiv:hep-th/0307084];
A.~Buchel,
  Phys.\ Rev.\  D {\bf 69}, 106004 (2004)
  [arXiv:hep-th/0312076];

\bibitem{Kahler}
    H. Kodama and K. Uzawa, JHEP 0603, 053 (2006) [arXiv:hep-th/0512104];
    K.~Koyama, K.~Koyama and F.~Arroja,
  Phys.\ Lett.\  B {\bf 641}, 81 (2006)
  [arXiv:hep-th/0607145].

\bibitem{frey} A.~R.~Frey and A.~Maharana,
  JHEP {\bf 0608}, 021 (2006)
  [arXiv:hep-th/0603233].


\bibitem{burges} C.~P.~Burgess, P.~G.~Camara, S.~P.~de Alwis, S.~B.~Giddings, A.~Maharana, F.~Quevedo and K.~Suruliz,
  arXiv:hep-th/0610255.

\bibitem{gt} G.~Torroba,
  JHEP {\bf 0702}, 061 (2007)
  [arXiv:hep-th/0611002].

\bibitem{gvw} S.~Gukov, C.~Vafa and E.~Witten,
  Nucl.\ Phys.\  B {\bf 584}, 69 (2000)
  [Erratum-ibid.\  B {\bf 608}, 477 (2001)]
  [arXiv:hep-th/9906070].
  
\bibitem{tv} T.~R.~Taylor and C.~Vafa,
  Phys.\ Lett.\  B {\bf 474}, 130 (2000)
  [arXiv:hep-th/9912152].

\bibitem{kklt}
S.~Kachru, R.~Kallosh, A.~Linde and S.~P.~Trivedi,
  Phys.\ Rev.\  D {\bf 68}, 046005 (2003)
  [arXiv:hep-th/0301240].

\bibitem{gray}J.~Gray and A.~Lukas,
  Phys.\ Rev.\  D {\bf 70}, 086003 (2004)
  [arXiv:hep-th/0309096].
  
\bibitem{pope}
P.~Hoxha, R.~R.~Martinez-Acosta and C.~N.~Pope,
  Class.\ Quant.\ Grav.\  {\bf 17}, 4207 (2000)
  [arXiv:hep-th/0005172];
C.~Pope, \emph{Kaluza-Klein Theory}, http://faculty.physics.tamu.edu/pope/ ihplec.pdf.  

\bibitem{SmallNumbers}
  S.~Dimopoulos, S.~Kachru, N.~Kaloper, A.~E.~Lawrence and E.~Silverstein,
  Int.\ J.\ Mod.\ Phys.\  A {\bf 19}, 2657 (2004)
  [arXiv:hep-th/0106128].

\bibitem{Mirage}
K.~Choi, A.~Falkowski, H.~P.~Nilles and M.~Olechowski,
  Nucl.\ Phys.\  B {\bf 718}, 113 (2005)
  [arXiv:hep-th/0503216];
  M.~Gabella, T.~Gherghetta and J.~Giedt,
  Phys.\ Rev.\  D {\bf 76}, 055001 (2007)
  [arXiv:0704.3571 [hep-ph]].

\bibitem{Sequestering}
	S.~Kachru, L.~McAllister and R.~Sundrum,
  JHEP {\bf 0710}, 013 (2007)
  [arXiv:hep-th/0703105];
K.~Choi,
  arXiv:0705.3330 [hep-th].
  
\bibitem{KKLMMT}
S.~Kachru, R.~Kallosh, A.~Linde, J.~M.~Maldacena, L.~P.~McAllister and S.~P.~Trivedi,
  JCAP {\bf 0310}, 013 (2003)
  [arXiv:hep-th/0308055].

\bibitem{DbraneInflation}
D.~Baumann, A.~Dymarsky, I.~R.~Klebanov and L.~McAllister,
  JCAP {\bf 0801}, 024 (2008)
  [arXiv:0706.0360 [hep-th]];
D.~Baumann, A.~Dymarsky, I.~R.~Klebanov, L.~McAllister and P.~J.~Steinhardt,
  Phys.\ Rev.\ Lett.\  {\bf 99}, 141601 (2007)
  [arXiv:0705.3837 [hep-th]];
A.~Krause and E.~Pajer,
  arXiv:0705.4682 [hep-th].
For a recent review, see
L.~McAllister and E.~Silverstein,
  Gen.\ Rel.\ Grav.\  {\bf 40}, 565 (2008)
  [arXiv:0710.2951 [hep-th]].

\bibitem{KofmanYi}
L.~Kofman and P.~Yi,
  Phys.\ Rev.\  D {\bf 72}, 106001 (2005)
  [arXiv:hep-th/0507257].

\bibitem{Chialva}
D.~Chialva, G.~Shiu and B.~Underwood,
  JHEP {\bf 0601}, 014 (2006)
  [arXiv:hep-th/0508229].

\bibitem{WarpedReheating}
N.~Barnaby, C.~P.~Burgess and J.~M.~Cline,
  JCAP {\bf 0504}, 007 (2005)
  [arXiv:hep-th/0412040];
A.~R.~Frey, A.~Mazumdar and R.~C.~Myers,
  Phys.\ Rev.\  D {\bf 73}, 026003 (2006)
  [arXiv:hep-th/0508139];
X.~Chen and S.~H.~Tye,
  JCAP {\bf 0606}, 011 (2006)
  [arXiv:hep-th/0602136];
 B.~v.~Harling, A.~Hebecker and T.~Noguchi,
  JHEP {\bf 0711}, 042 (2007)
  [arXiv:0705.3648 [hep-th]].

\bibitem{tye} H.~Firouzjahi and S.~H.~Tye,
  JHEP {\bf 0601}, 136 (2006)
  [arXiv:hep-th/0512076].

\bibitem{Kofman2}
J.~F.~Dufaux, L.~Kofman and M.~Peloso,
  arXiv:0802.2958 [hep-th].

\bibitem{WarpedLHC}
G.~Shiu, B.~Underwood, K.~M.~Zurek and D.~G.~E.~Walker,
  Phys.\ Rev.\ Lett.\  {\bf 100}, 031601 (2008)
  [arXiv:0705.4097 [hep-ph]].

\bibitem{Grimm} T.~W.~Grimm and J.~Louis,
  Nucl.\ Phys.\  B {\bf 699}, 387 (2004)
  [arXiv:hep-th/0403067].
  T.~W.~Grimm,
  Fortsch.\ Phys.\  {\bf 53}, 1179 (2005)
  [arXiv:hep-th/0507153].
    
\bibitem{wald} R.~M.~Wald, \emph{General Relativity}, University of
Chicago Press, 1984; chapter 7.

\bibitem{RadionDimReduct}
C.~Csaki, M.~L.~Graesser and G.~D.~Kribs,
  Phys.\ Rev.\  D {\bf 63}, 065002 (2001)
  [arXiv:hep-th/0008151].

\bibitem{candelas} P.~Candelas and X.~de la Ossa,
  Nucl.\ Phys.\  B {\bf 355}, 455 (1991).

\bibitem{ghy} J.~W.~.~York,
  Phys.\ Rev.\ Lett.\  {\bf 28} (1972) 1082.
  G.~W.~Gibbons and S.~W.~Hawking,
  Phys.\ Rev.\  D {\bf 15}, 2752 (1977).

\bibitem{dterm}
D.~E.~Diaconescu, M.~R.~Douglas and J.~Gomis,
  JHEP {\bf 9802}, 013 (1998)
  [arXiv:hep-th/9712230].
M.~Aganagic and C.~Beem,
  Nucl.\ Phys.\  B {\bf 796}, 44 (2008)
  [arXiv:0711.0385 [hep-th]].

\bibitem{weinberg} S.~Weinberg, \emph{The Quantum Theory of Fields}, volume III, Cambridge University Press, 2000.

\bibitem{duff} A.~Salam and J.~A.~Strathdee,
  Annals Phys.\  {\bf 141}, 316 (1982).
L.~Dolan and M.~J.~Duff,
  Phys.\ Rev.\ Lett.\  {\bf 52}, 14 (1984).

\bibitem{cho} Y.~M.~Cho and S.~W.~Zoh,
  Phys.\ Rev.\  D {\bf 46}, 2290 (1992).

\bibitem{msugra} W.~D.~I.~Linch, M.~A.~Luty and J.~Phillips,
  Phys.\ Rev.\  D {\bf 68}, 025008 (2003)
  [arXiv:hep-th/0209060].
  T.~Gregoire, M.~D.~Schwartz and Y.~Shadmi,
  JHEP {\bf 0407}, 029 (2004)
  [arXiv:hep-th/0403224].
\bibitem{DT} M.~R.~Douglas and G.~Torroba, to appear.

\bibitem{aharony} O.~Aharony, Y.~E.~Antebi and M.~Berkooz,
  Phys.\ Rev.\  D {\bf 72}, 106009 (2005)
  [arXiv:hep-th/0508080].

\end{thebibliography}
\end{document}